\documentclass[10pt,journal,compsoc]{IEEEtran}
\usepackage{graphicx}
\usepackage{textcomp}
\usepackage{subfigure}
\usepackage{makecell}
\usepackage{balance}
\usepackage{bm}
\usepackage{amsfonts}
\usepackage{multirow}
\usepackage{algorithm}
\usepackage{algorithmic}
\usepackage{amsmath}
\usepackage{amsthm}
\usepackage{amssymb,color}
\usepackage{float}
\usepackage{graphicx}
\usepackage{caption}
\usepackage{ragged2e}

\newtheorem{assumption}{Assumption}

\newtheorem{theorem}{Theorem}
\ifCLASSOPTIONcompsoc
  \usepackage[nocompress]{cite}
\else
  \usepackage{cite}
\fi
\ifCLASSINFOpdf
\else
\fi
\begin{document}
%
\title{EI-MTD: Moving Target Defense for Edge Intelligence against Adversarial Attacks}
%
%
%
%

\author{Yaguan Qian, Qiqi Shao, Jiamin Wang, Xiang Ling, Yankai Guo,\\ Zhaoquan Gu*, Bin Wang*, and Chunming Wu* 

\IEEEcompsocitemizethanks{
\IEEEcompsocthanksitem Yaguan Qian, Qiqi Shao, Jiamin Wang, and Yankai Guo are with School of Big-Data Science, Zhejiang University of Science and Technology, Hangzhou {\rm 310023}, China.\protect

\IEEEcompsocthanksitem Xiang Ling and Chunming Wu are with College of Computer Science and Technology, Zhejiang University, Hangzhou {\rm 310058}, China.

\IEEEcompsocthanksitem Zhaoquan Gu is with Cyberspace Institute of Advanced Technology (CIAT), Guangzhou University, Guangzhou {\rm 510006}, China.

\IEEEcompsocthanksitem Bin Wang is with Network and Information Security Laboratory of Hangzhou Hikvision Digital Technology Co., Ltd. Hang Zhou {\rm 310052}, China.}



}

\IEEEtitleabstractindextext{%
\begin{abstract}\justifying
Edge intelligence has played an important role in constructing smart cities, but the vulnerability of edge nodes to adversarial attacks becomes an urgent problem. The so-called adversarial example can fool a deep learning model on the edge node for misclassification. Due to the transferability property of adversarial examples, the adversary can easily fool a black-box model by a local substitute model. The edge nodes have limited resources, which cannot afford a complicated defense mechanism as that on the cloud data center. To address the challenge, we propose a dynamic defense mechanism, namely EI-MTD. The mechanism first obtains robust member models of small size through differential knowledge distillation from a complicated teacher mode on the cloud data center. Then, a dynamic scheduling policy, which builds on a Bayesian Stackelberg game, is applied to the choice of a target model for service. This dynamic defense mechanism can prohibit the adversary from selecting an optimal substitute model for black-box attacks. We also conduct extensive experiments to evaluate the proposed mechanism, and the results show that EI-MTD could protect edge intelligence effectively against adversarial attacks in black-box settings.  

\end{abstract}

}

\maketitle

\IEEEdisplaynontitleabstractindextext

%
\IEEEpeerreviewmaketitle

\IEEEraisesectionheading{\section{introduction}\label{sec:introduction}}

\IEEEPARstart{A}{rtificial} intelligence (AI) based on deep learning has been successfully applied in various fields, ranging from facial recognition \cite{a1}, natural language processing \cite{[a2]}, to computer vision \cite{[a3]}. With the vigorous development of AI, people are increasingly relying on the convenient services provided by intelligent life, hoping to enjoy intelligent services anytime and anywhere. In the past few years, the advancement of edge computing has moved from theory to application and various applications have been developed to improve our daily lives. The maturity of deep learning techniques and edge computing systems \cite{[a5]}, \cite{[a6]}, and the increasing demand for intelligent life \cite{[a7]}, \cite{[a8]} promote the development and realization of \textit{edge intelligence} (EI). The current EI is based on deep learning models, i.e., deep neural networks (DNNs) that are deployed to devices on edge of networks (such as an intelligent camera of monitoring systems) to realize real-time applications such as target recognition and anomaly detection.

At present, the security of EI becomes a wide concerned problem. Most previous work focuses on data privacy of EI, however, it does not pay enough attention to adversarial attacks. The previous work shows that DNNs are extremely vulnerable to \textit{adversarial examples} \cite{[a14]}. In image classification scenarios, an adversarial example is an input image added by some well-designed tiny perturbations to fool a DNN that is referenced as a \textit{target model}. A special property of adversarial examples is transferability, implying that adversarial examples that successfully fool one model can also be misclassified by other models with a high probability \cite{[a64]}. Theoretically, with this transferability property, the adversary crafts adversarial examples on a local \textit{substitute model} without knowing any information about the target model \cite{[a45]}, which is one kind of \textit{black-box attacks}. Practically, the adversary attempts to find a proper substitute model close to the target model by repeated queries \cite{[a73]}, \cite{[a81]}, which can finally obtain a higher attack success rate approximate to \textit{white-box attacks} that the adversary fully knew the information on the target model \cite{[a80]}.

Due to the limited resources on edge nodes including edge devices and edge servers, model compression is considered as an effective way to reduce the model size for EI \cite{[a10]}, \cite{[a13]}. However, adversarial attacks make the situation more challenging since Madry et. al \cite{[a21]} observed that the robustness of DNNs is positively correlated to the model size. L. Wu et. al \cite{[a80]} revealed that if the target model has a smaller model size, an adversarial example is easier to achieve a higher transfer rate. Thus, compressed models on edge nodes become more vulnerable to adversarial examples. Besides, although many methods are proposed to defend against adversarial attacks \cite{[a21]}, \cite{[a18]}, \cite{[a19]}, \cite{[a20]}, they merely work well on a cloud data center with abundant GPUs, which is not applicable to edge nodes. Therefore, the limitation of resources aggravates the vulnerability of EI in the face of adversarial attacks.

Based on the above statement, we summarized the challenges for EI to address in adversarial settings as follows: (1) how to prevent the adversary from finding a proper substitute model, (2) how to reduce the transferability of adversarial attacks without compromising accuracy, and (3) how to defend against adversarial examples for edge nodes with limited resources. For the first challenge, we change the setting from a static target model to a dynamic setting that randomly schedules a model for classification services. Since the adversary does not know the true model served for them, they cannot estimate which candidate substitute model approximates the target model. For the second challenge, we attempt to increase the diversity of different models deployed on edge nodes. We use the gradient of a loss function as a basis of the difference measure since the current attacks mainly use gradients to craft adversarial examples. For the third challenge, we utilize transfer learning to distill the knowledge from a robust teacher model with a large size on the cloud data center to student models with small size. The benefit of this approach is that the classification knowledge and robustness are transferred as well as the size of models is compressed.These aforementioned techniques are not independent but supplement each other. They are integrated into our proposed defense framework represented as EI-MTD (Edge Intelligence with Moving Target Defense).

%
 Although there are several MTDs proposed for adversarial attacks, they are not aimed at the defense of EI. Sailik et. al in \cite{[a23]} proposed an MTDeep to defend DNN models, while our approach is different from them in two aspects. First, MTDeep merely focuses on cloud data centers, which does not suitable for EI with limited resources. Second, for obtaining differential configurations, their member models include CNNs, MLPs, and HRNNs, while we use the same type of DNNs. The benefit of our method is that all member models can perfectly serve some specific tasks such as facial recognition. Furthermore, we use differential knowledge distillation to realize the diversity of member models is more efficient than MTDeep. Similar to Sailik, Roy et. al in \cite{[a24]} proposed MTD against adversarial attacks in a cloud computing scenario which is also not suitable for EI. In addition, their member models are heterogeneous, including SVMs, logistic regression and CNNs which cannot fully utilize the advanced deep learning techniques. Song et. al in \cite{[a25]} designed an fMTD for embedded deep visual sensing systems against adversarial examples. Since their fMTD is only limited to a single device, it cannot be well extended to EI with many nodes. Moreover, they “fork” member models by retraining the perturbed base model, which cannot effectively guarantee the diversity of member models to overcome transferability.

The main innovations and contributions of our work can be summarized as follows:

\begin{enumerate}
  \item [(1)] To the best of our knowledge, we are the first to propose a defense mechanism, namely EI-MTD for EI against adversarial attacks. We propose a dynamic scheduling policy by means of the Bayesian Stackelberg game to defend against black-box adversarial attacks. This method perfectly combines the inference architecture of EI, i.e., a deep learning model independently performs inference on an edge node, to dynamic execution. This dynamic scheduling mechanism is completely transparent to users and easy to be implemented without damaging accuracy.

  \item [(2)] To prevent transferability, we propose differential knowledge distillation to increase the diversity of member models on edge nodes. Different from the Hinton’s knowledge distillation for a single model, we distillate $K$ student models simultaneously with a common loss function combined with a differential regularization term. In addition, this method overcomes the limitation of resources due to its effect on transferring robust knowledge and compressing models.

  \item [(3)] We built an EI simulation platform with several GPU servers, PC, and Raspberry pi to evaluate our EI-MTD. A real image dataset ILSVRC2012 \cite{[a75]} is used for our experiments. All the adversarial examples are generated with typical attack methods: FGSM \cite{[a15]}, PGD \cite{[a21]},I-FGSM \cite{[a16]}, and M-DI$^{2}$-FGSM \cite{[a17]}. Extensive experimental results show that EI-MTD can increase $\thicksim$43\% accuracy against black-box adversarial attacks generated with the state-of-the-art M-DI$^{2}$-FGSM.
     
\end{enumerate}

The rest of the paper is organized as follows. In Section 2, we introduce the preliminary of DNNs, adversarial examples, knowledge distillation, and the Bayesian Stackelberg game, etc. In Section 3, the assumptions are presented and the problems we need to address are defined. In Section 4, the key techniques including differential knowledge distillation and scheduling policy in EI-MTD are described in detail. In Section 5, extensive experimental results are discussed and some interesting conclusions are obtained. In Section 6, related work about adversarial attacks and countermeasures is presented. In Section 7, we make a conclusion and discuss some future work.

\section{PRELIMINARIES}\label{section_introduction}\label{Sec:Def1}

In this section, we will formally introduce the concepts of DNNs and adversarial examples. Besides, we introduce in brief the principles of knowledge distillation and the Bayesian Stackelberg game that are the basis of our proposed EI-MTD.

\subsection{DNNs and Adversarial Examples}
A DNN is often denoted as a function: $F(x, \theta): \mathbb{R}^{d} \rightarrow \mathbb{R}^{L}$ , where $x \in \mathbb{R}^{d}$ is an input, $\theta$ denotes model parameters, and $L$ is the number of classes. In this paper, we assume a DNN contains a softmax layer that is defined as follows:
\begin{equation}
\text {Softmax}(z)_{i}=\frac{\exp \left(z_{i}\right)}{ \sum\limits_{i=1}^{L} \exp \left(z_{i}\right)}, i \in\{1, \ldots, L\}
\end{equation}
where $z=\left(z_{1}, z_{2}, \ldots, z_{L}\right)$ is the output of the last hidden layer. Therefore, a DNN can also be denoted as $F(x)=\operatorname{Softmax}(z)$. With the above definition, the class prediction is obtained by $\hat{y}=\operatorname{argmax}_{i \in\{1, \ldots,L\}} F(x)_{i}$, where $F(x)_{i}$  is the confidence score of the $i$-th class.

An adversarial example $x_{adv}$ is a clean image $x$ added by a well-designed tiny perturbation $r$ that is almost imperceptible to human eyes but can easily fool a DNN. In general, the process to find the perturbation is modeled as the following optimization problem:
\begin{equation}\begin{array}{l}
\arg \min _{r}\|r\|_{p}\\
\text {s.t.\quad arg} \max \limits_{i \in\{1, \ldots,L\}} F\left(x_{adv}\right) \neq y
\end{array}
\end{equation} 
where $x_{adv}=x+r$, $\|\cdot\|_{p}$ is $p$-norm ($p=1,2,$ or $\infty$) and $y$ is the ground-truth label of $x$.

\subsection{Knowledge Distillation}
There have been a number of attempts to transfer knowledge between varying-capacity network models. Hinton et.al \cite{[a27]} is the first to distill knowledge from a large pre-trained teacher model to improve a small target network that is also called a student model. In the context of EI-MTD, a \textit{student model} is equivalent to a member model for scheduling. To perform distillation, a new softmax function is defined for the pretrained teacher model $F_{teacher}(\theta)$:
\begin{equation}
\tilde{y}(x)=\left[\frac{e^{z_{i}(x) / T}}{\sum \limits_{l=0}^{L-1} e^{z_{i}(x) / T}}\right]_{i \in 1, \ldots, L}
\end{equation} 
where $T$ is temperature and $\tilde{y}(x)$ is the soft label corresponding to the hard label $y$. Notice that $\tilde{y}(x)$ is substantially a confidence vector. The so-called knowledge distillation is to train a student model $F_{\text {student}}\left(\theta_{s}\right)$ with soft labels by minimizing the following loss function:\begin{equation}L=\beta T^{2} J\left(x, \tilde{y} ; \theta_{s}\right)+(1-\beta) J\left(x, y ; \theta_{s}\right)\end{equation} where $J$ is a cross-entropy loss, $\theta_{s}$ is the parameters of the student model, and $\beta$ is a hyperparameter that controls the relative importance between soft labels and hard labels. Instead of training only with hard a label $y$ traditionally, the student model trained with a soft label $\tilde{y}(x)$ can achieve higher accuracy. In this paper, we use knowledge distillation to transfer robustness knowledge from a robust model on the data center to member models on edge nodes and compress models as well.

\subsection{Bayesian Stackelberg Game}
A Stackelberg game is a non-cooperative, hierarchical decision-making game, which is played between two players called a leader and a follower respectively. Let us denote the Stackelberg game as a six-tuple $G=\left(L, F, S_{L}, S_{F}, R_{L}, R_{F}\right)$, where $L$ represents a leader, $F$ represents a follower, $S_{L}$ is the action space of a leader, $S_{F}$ is the action space of a follower, $R_{L}$ and $R_{F}$ is a payoff function of a leader and a follower respectively. A payoff function is defined on a combination of actions:
\begin{equation}R_{i}:\left[S_{L}\right] \times\left[S_{F}\right] \rightarrow \mathbb{R}\end{equation} 
where $\left[S_{i}\right]$ denotes the index set of the action space and $i=L,F$. In the Stackelberg game, a leader commits a mixed strategy $s$ first, and then a follower $F$ optimizes its payoff with a pure strategy $q$ according to the action chosen by the leader. Here, a pure strategy means only one action to be chosen, while a mixed strategy is a distribution defined on the action space, that is, every action is chosen with a probability.

In the field of security, it usually considers the defender as a leader and the follower in general contains multiple types. In such a case, the Stackelberg game needs to be extended to multiple follower types, which is called the Bayesian Stackelberg game denoted as $G_{Bayes}=\left(L, F^{(c)}, S_{L}, S_{F^{(c)}}, R_{L^{(c)}}, R_{F^{(c)}}, p^{(c)}\right)$, $c \in 1, \ldots, C$. Suppose a follower contains $C$ types, each type of follower $F^{(c)}$ has its own strategy set $S_{F^{(c)}}$ and payoff function $R_{F^{(c)}}$, and $p^{(c)}$ denotes the probability of $F^{(c)}$ occurrence. We can define the payoff function of the follower as follows:

Since the leader does not know the type of the follower $F^{(c)}$, but merely knows the probability distribution $p^{(c)}$ of follower types, a Bayesian Stackelberg game belongs to an incomplete information game. Our goal is to ﬁnd the optimal mixed strategy for the leader to commit to; given that the follower may know this mixed strategy when choosing its strategy. Finally, for obtaining the optimal mixed strategy, we solve the following problem:
\begin{equation}\begin{array}{l}
\max\limits_{s, q, v} \sum\limits_{i \in\left[S_{L}\right]}\sum\limits_{c \in C} \sum\limits_{j \in\left[S_{F}\right]} p^{(c)} R_{t^{(c)}}(i, j) s_{i} q_{j}^{(c)} \\
\text {s.t. }
 0 \leq\left(v^{(c)}-\sum\limits_{i \in\left[S_{L}\right]} R_{F^{(c)}}(i, j) s_{i}\right) \leq\left(1-q_{j}^{(c)}\right)N\\
\qquad\qquad\qquad\qquad \sum\limits_{i \in\left[S_{L}\right]} s_{i}=1\\
\qquad\qquad\qquad\qquad\sum\limits_{j \in\left[S_{F}\right]} q_{j}^{(c)}=1 \\
\qquad\qquad\qquad\qquad\mathrm{s}_{i} \in[0 ,1]\\
\qquad\qquad\qquad\qquad q_{j}^{(c)} \in\{0,1\} \\
\qquad\qquad\qquad\qquad v^{(c)} \in \mathbb{R}
\end{array}\end{equation}
Problem (6) is a mixed-integer quadratic programming (MIQP) that can be solved by DOBSS \cite{[a26]}, and the solution is termed Bayes-Nash equilibrium containing the leader’s optimal mixed strategy $s$, the optimal response strategy $q^{(c)}$ and the payoff of the follower $v^{(c)}$.

\section{PROBLEM SETUP}\label{Sec:background}

\subsection{Assumptions}

\begin{assumption}
We first assume the whole DNN model runs on an edge node, which is not divided into several parts distributed to other edge nodes or cloud servers.
\end{assumption}
The inference architectures of DNN-based EI can be classified into edge-based, device-based, edge-device, and edge-cloud modes \cite{[a28]}. We assume a teacher model completed its \textit{adversarial training} \cite{[a15]} on the cloud data center and then compressed into the student models by knowledge distillation. A student model is finally deployed on an edge device or an edge server to independently perform inference. Accordingly, the proposed EI-MTD supports both edge-based and device-based modes, which facilitate the scheduling of EI-MTD. Under this assumption, we further discuss the adversary’s attack strategy and the problem to be addressed.

\begin{assumption}
We assume our defenses against image adversarial examples in black-box attacks settings. 
\end{assumption}
Since image classification is the most common application of EI, and image adversarial examples are widely studied so far, EI-MTD is mainly designed to defend image adversarial examples. In general, the adversary crafting adversarial examples needs clean images and the information of a target model as shown in Eq. (2). According to the adversary’s ability to obtain the information of a target model, it can be categorized into (1) \textit{black-box attacks} that the adversary knows nothing about the target DNN expect for prediction labels or confidence vectors and (2) \textit{white-box attacks} that the adversary has the full information of the target DNN, including training data, model parameters, topology structures, and training methods, etc. In real life, black-box attacks are more practice for the adversary \cite{[a45]}.

\begin{assumption}
The adversary’s goal is to achieve a higher attack success rate under the black-box scenario. Hence, we assume the adversary’s strategy is to selected an optimal substitute model by a repeated query like \cite{[a45]}, \cite{[a73]}. 
\end{assumption}
Suppose there are $K$ edge nodes $E=\left\{e_{i}\right\}_{i=1}^{K}$ and $K$ target models $T=\left\{F_{t}\left(\theta^{(i)}\right)\right\}_{i=1}^{K}$, and each node has a target model, e.g., $F_{t}\left(\theta^{(k)}\right)$on the edge node $e_{k}$. We formalize the adversary’s strategy in EI settings as follows. The adversary has $M$ local substitute models $U=\left\{F_{a}\left(\theta^{(i)}\right)\right\}_{i=1}^{M}$ for choice to generate adversarial examples as shown in the left of  Fig. \ref{fig1}. Assume the adversary’s attack target is the model $F_{t}\left(\theta^{(k)}\right) \in T$ on the node $e_{k}$. The adversary queries $F_{t}\left(\theta^{(k)}\right)$ with $M$ adversarial examples $x_{adv}^{(i)}=x+r^{(i)}, i=1, \ldots, M$ every time. Notice that crafting $M$ adversarial examples with the same clean image $x$ in a query. Let $c^{(i)}=\mathbb{I}\left(F_{t}\left(x_{a d v}^{(i)}, \theta^{(k)}\right) \neq y\right)$ where $\mathbb{I}(\cdot)$ is an indicator function and $y$ is the ground-truth label of $x$. After $n$ queries, the adversary obtains an optimal substitute model $F_{a}\left(\theta^{\left(i^{*}\right)}\right)=\arg \max _{i}\left\{\frac{1}{n} \sum c^{(i)}\right\}_{i=1}^{M}$. With this optimal substitute model, the adversary can continue to conduct a black-box attack on the target model with a higher success rate.

\subsection{Problems Statement}
Based on the above-mentioned assumptions, we attempt to address the following problems related to the countermeasure for EI against black-box attacks:

\begin{itemize}
  \item [(1)] How to prevent the adversary from finding an optimal substitute model? The reason for the adversary’s success is that classification services provided by EI are static, i.e., all the queries to $e^{k}$ are responded by $F_{t}(\theta^{(k)})$, which enables the adversary to obtain consistent information by repeated queries. We propose a dynamic strategy to confuse the adversary via randomly switching models.

  \item [(2)] How to reduce the transferability of adversarial attacks without compromising accuracy? Even if we perform a dynamic strategy, it still has no truly effective if each target model in $T=\left\{F_{t}\left(\theta^{(i)}\right)\right\}_{i=1}^{N}$ is identical. Despite this special case, the adversarial examples yet succeed across different target models. This transferability will degrade the effectiveness of our dynamic strategy. Hence a well-designed method to increase the diversity of member models is necessary.

  \item [(3)] How to defend against adversarial examples with limited resources on edge nodes? Since we assume the whole DNN model runs on an edge node, the size of the model must be enough smaller to meet the limited resources, especially the memory requirement. However, a model with a smaller size is in general more vulnerable than a model with a larger size \cite{[a21]}. We hope to obtain a smaller model, at the same time to ensure its accuracy and robustness.
     
\end{itemize}

\section{METHODOLOGY}
\begin{figure}
	\centering  
	\includegraphics[width=8.8cm,height=3.8cm]{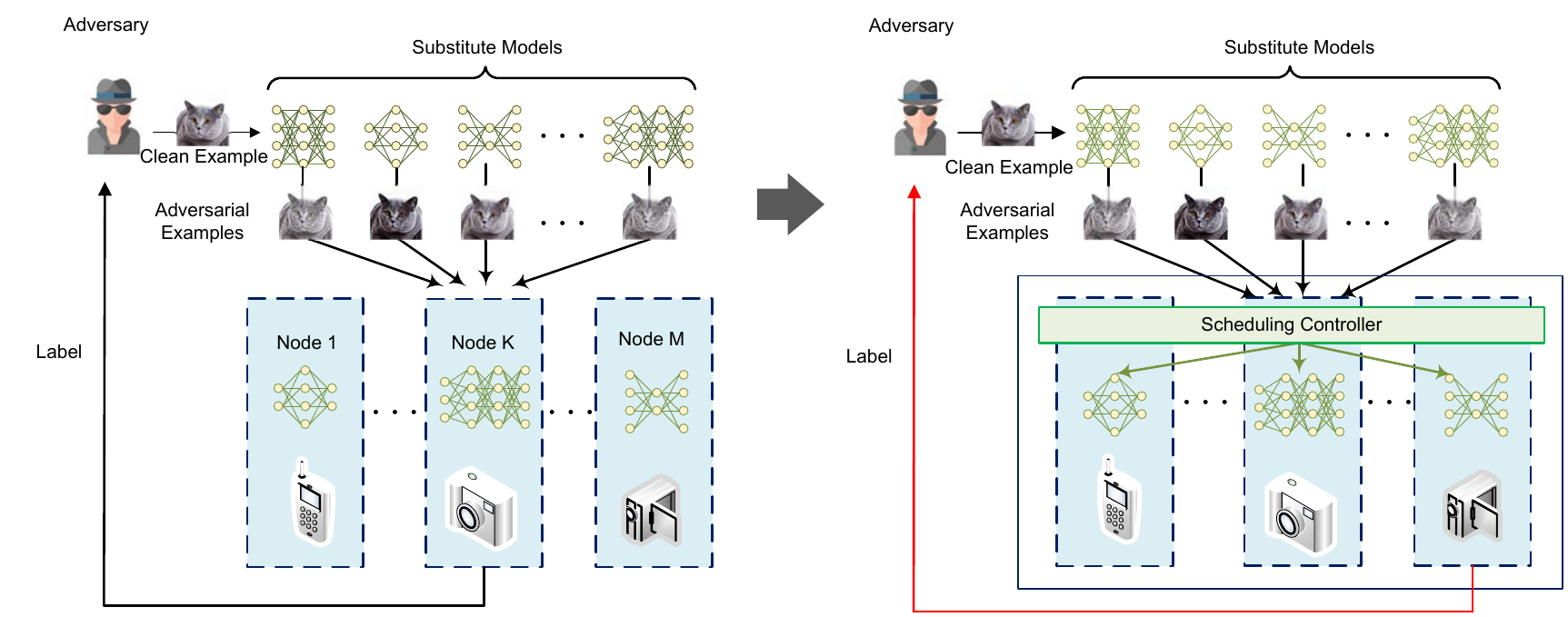}  
	\caption{Static target models vs. moving target models. The left is the typical device-based interference architecture with static service. The adversary attack Node K with a “cat” adversarial example and he has known the model on Node K performs classification. The right is our dynamic scheduling scheme. Although the adversary attempts to attack Node K, he is not aware of the actual performing model on which nodes.}  
	\label{fig1}   
\end{figure}

In this section, we will first describe our method at a high level. To address the problems raised in Section 3.2, we proposed EI-MTD that includes three key techniques: adversarial training, differential knowledge distillation, and service dynamic scheduling. We use adversarial training to obtain a robust teacher model on the cloud data center. Then we use transfer learning to distill the robust knowledge from the teacher model to student models with small size for limited resources. At the same time, different from Hinton’s knowledge distillation \cite{[a27]}, we add a differential regularization term to obtain a diversity of distilled models, which can effectively prohibit the transferability. These student models, also called member models under the MTD context are further used for scheduling in our service dynamic scheduling scheme. Benefit from the obtained diversity, our dynamic scheduling can perfectly confuse the adversary to find an optimal substitute model as shown in the right of Fig. \ref{fig1}.

\begin{itemize}
  \item [(1)] \textbf{Adversarial training for a teacher model.} Suppose we have a training dataset $D=\left\{\left(x_{i}, y_{i}\right)\right\}_{i=1}^{N}$ and a teacher model $F_{t}\left(\theta_{t}\right)$. Previous work shows that a larger network can achieve more robustness by adversarial training \cite{[a21]}. Hence, we choose the network such as ResNet-101 \cite{[a29]} with 101 layers as our teacher model. However, adversarial training for a deeper network needs more computation resources and time than standard training. As a result, we perform adversarial training on the cloud data center, and use the recently proposed “FAST” adversarial training method to accelerate the process. Due to the limitation of paper length, we do not describe the training detail in this paper, which can be referred to \cite{[a20]}.

  \item [(2)] \textbf{Differential knowledge distillation for student models.} We first use Eq. (3) to obtain a soft label $\tilde{y}_{i}$ of $x_{i}$ from the teacher model $F_{t}\left(\theta_{t}\right)$ with an appropriate distillation temperature $T$ and then create a new training dataset $\tilde{D}=\left\{\left(x_{i}, \tilde{y}_{i}\right)\right\}_{i=1}^{N}$. The essence of knowledge distillation is to train a model with soft labels. To obtain the diversity of student models, we define a new loss function $L=\sum T^{2} J / K+\lambda \cdot C S_{\text {coherene}}$ with a regularization term $C S_{\text {coherence }}$ that will be discussed in detail in Section 4.1. Thus, we simultaneously train all the student models in $V=\left\{F_{t}\left(\theta^{(t)}\right)\right\}_{i=1}^{K}$ with dataset $\tilde{D}$ to minimize the common loss function $L$. Notice student model, member model, and target model are referred to the same object in this paper and they have a certain name under a specific context.

  \item [(3)] \textbf{Dynamic Service Scheduling for member models.} After differential knowledge distillation, the student models are deployed to edge nodes. Remind that each edge node, include edge device and edge server, has only one student model. Among them, an edge server is assigned to act as a scheduling controller. All the student models, i.e., member models are registered at the scheduling controller. When a user (including an adversary) inputs an image through an edge device, e.g., a smartphone, for classification service, the edge device first uploads it to the scheduling controller instead of directly processing on its local model. The scheduling controller selects an edge node, to be more precise, the model on it to perform classification tasks. Therefore, the adversary cannot be aware of which edge node to provide service ultimately. Obviously, the scheduling policy plays a key role in EI-MTD. We use the Bayesian Stackelberg game to provide an optimal choice that will describe in Section 4.3 in detail.

 \end{itemize}

\subsection{Measure of Diversity}
As above mentioned that the diversity of models play an important role in the effectiveness of dynamic scheduling. To this end, how to properly measure the diversity is a non-trivial problem. Inspired by the fact that state-of-the-art adversarial attacks leverage gradients with respect to input examples as the direction of perturbation, we use gradient alignment as a diversity metric instead of differential immunity for differential knowledge distillation.

Suppose there are two member models $F_{t}^{(1)}$ and $F_{t}^{(2)} \in \Omega$, and a substitute model $F_{a} \in U$ selected by the adversary. Let $\nabla_{x} J_{t}^{(i)}$ denote the gradient of the loss function of the member model $F_{t}^{(i)}$ with respect to an input $x$. If the angel between $\nabla_{x} J_{t}^{(1)}$ and $\nabla_{x} J_{t}^{(2)}$ is small enough, then an adversarial example that successfully fools $F_{t}^{(1)}$ can also fool $F_{t}^{(2)}$ with a high probability. Accordingly, the angel between $\nabla_{x} J_{t}^{(1)}$ and $\nabla_{x} J_{t}^{(2)}$ are related to the difference between models and we introduce cosine similarity (CS) to measure the difference of models for differential knowledge distillation: \begin{equation}C S\left(\nabla_{x} J_{t}^{(1)}, \nabla_{x} J_{t}^{(2)}\right)=\frac{<\nabla_{x} J_{t}^{(1)}, \nabla_{x} J_{t}^{(2)}>}{\left|\nabla_{x} J_{t}^{(1)}\right| \cdot\left|\nabla_{x} J_{t}^{(2)}\right|}\end{equation} where $<\nabla_{x} J_{t}^{(1)}, \nabla_{x} J_{t}^{(2)}>$ is the inner product of $\nabla_{x} J_{t}^{(1)}$ and $\nabla_{x} J_{t}^{(2)}$. If $CS\left(\nabla_{x} J_{t}^{(1)}, \nabla_{x} J_{t}^{(2)}\right)=-1$, it means $\nabla_{x} J_{t}^{(1)}$ and $\nabla_{x} J_{t}^{(2)}$ are completely misaligned, that is, an adversarial example that can fool $F_{t}^{(1)}$ cannot fool $F_{t}^{(2)}$.

\subsection{Differential Knowledge Distillation}
In this section, we further apply cosine similarity to knowledge distillation to obtain differential member models. Since the cosine similarity is computed with two gradients and our EI-MTD includes $K(K \geq 2)$ it needs to extend Eq. (7) to the situation with $K$ gradients. We define the maximum over all pairwise cosine similarity as EI-MTD diversity measure:
\begin{equation}\begin{split}
CS_{\text {coherence}}=\max _{a, b \in\{1, \ldots, K\} \wedge a \neq b} CS & \left(\nabla_{x} J_{t}^{(a)}\left(x, \tilde{y} ; \theta^{(a)}\right),\right.\\
&\left.\nabla_{x} J_{t}^{(b)}\left(x, \tilde{y} ; \theta^{(b)}\right)\right)
\end{split}\end{equation}
where $J_{a}$ and $J_{b}$ are the loss function of member models $F_{s}^{(a)}$ and $F_{s}^{(b)}$ respectively, and $\tilde y$ is the soft label of $x$ obtained from the teacher model. Since Eq. (8) is not a smooth function, it cannot be solved with a gradient descent method. We use the LogSumExp function to smoothly approximate $CS_{coherence}$ as follows:

\begin{equation}\begin{array}{l}
CS_{\text{coherence}}\approx\log( \sum\limits_{\mathrm 1<a<b<K}\exp(CS(\nabla_{x} J_{t}^{(a)}(x, \tilde{y};\theta^{(a)})\\
\qquad\qquad\qquad\qquad\qquad\qquad\qquad,\nabla_{x} J^{(b)}(x, \tilde{y} ; \theta^{(b)} ) ) ) )
\end{array}\end{equation}

A small $CS_{coherence}$ implies a large diversity between member models in EI-MTD. Remind that the member model is distilled from the teacher model of the cloud data center. At the same time, we need to guarantee the diversity between member models, hence we add $CS_{coherence}$ to the process of knowledge distillation as a regularization term. We define a new loss function for distillation as follows: \begin{equation}L=\frac{1}{K} \sum_{i=1}^{K} T^{2} J_{t}^{(i)}\left(x, \tilde{y} ; \theta^{(i)}\right)+\lambda \cdot C S_{\text {coherence}}\end{equation} where $\lambda$ is a regularization coefficient to control the importance given to $CS_{coherence}$ during training. We set $\beta=1$ in Eq. (4) to enable student models to learn enough knowledge from teacher models, that is, we only use soft-label examples to train student models. Our proposed differential knowledge distillation is shown in Algorithm 1.


\begin{algorithm}
	\caption{Difference knowledge distillation}
     
    {\textbf{Input:} The $K$ student models $\left\{F\left(\theta^{(k)}\right)\right\}_{k=1}^{K}$; the training dataset $D=\left\{\left(x_{i}, y_{i}\right)\right\}_{i=1}^{N}$; the distillation temperature $T$; the learning rate: $\left\{\eta^{(k)}\right\}_{k=1}^{K}$}\\
    {\textbf{Output:} The parameters $\theta^{(k)}$ of student models}

\begin{algorithmic}[1]

		\STATE initialize each $\theta^{(k)}$ as $\theta_{0}^{(k)}$ 
		\STATE $\tilde{\mathrm{y}}_{i}=F_{t}\left(\operatorname{softmax}\left(\mathrm{Z}\left(x_{i}\right), T\right)\right)$, $\tilde{D}=\left\{\left(x_{i}, \tilde{\mathrm{y}}_{i}\right)\right\}_{i=1}^{N}$
		\STATE \textbf{for}  $m=1,...,epochs$ \textbf{do}\\
           \STATE \qquad $L_{m}=\frac{1}{K} \sum_{i=1}^{K} T^{2} J_{t}^{(i)}\left(x, \tilde{y} ; \theta^{(i)}\right)+\lambda \cdot C S_{\text {coherence}}$\\
		\STATE  \qquad \textbf{for}  $k=1,...,K$ \textbf{do}\\
		\STATE  \qquad\qquad $\theta_{m+1}^{(k)} \leftarrow \theta_{m}^{(k)}-\eta^{(k)} \nabla_{\theta_{m}^{(k)}} L_{m}$
		\STATE    \qquad \textbf{end for}\\
		\STATE  \textbf{end for}\\
		
\end{algorithmic}
\end{algorithm}

\subsection{Service Scheduling Policies}
\begin{figure}
	\centering  
	\includegraphics[width=6.5cm,height=6cm]{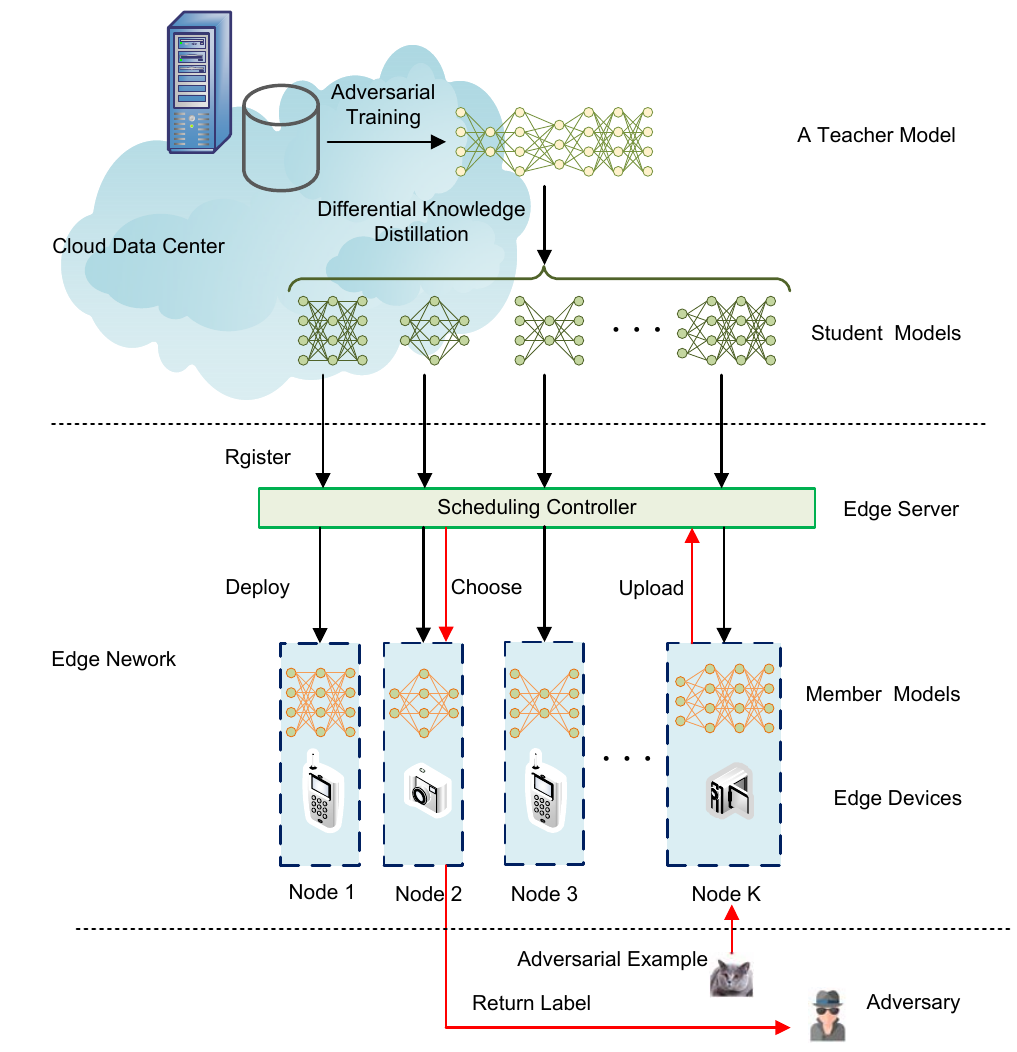}  
	\caption{The framework of EI-MTD, where the black lines represent the process of member models deployed to edge nodes, and the red lines represent the process of an adversarial image by EI-MTD.}  
	\label{fig2}   
\end{figure}

After differential knowledge distillation, the student models are deployed to edge nodes as shown in Fig. \ref{fig2}. These models are prepared to perform image classification tasks as following steps. First, the user inputs an image at an edge device. The edge device accepts the image but does not perform classification directly on its own model. Instead, it forwards the image to the scheduling controller. When the scheduling controller receives the request, it selects a registered edge node by a scheduling policy and sends the image to the selected node. In this section, we will describe the scheduling policy in detail.

We use a Bayesian Stackelberg game to model our scheduling policy. Recall that the adversary’s strategy is to select an optimal substitute model to achieve a higher attack success rate while the defender’s strategy is to randomly select an edge node for service to confuse the adversary. Nevertheless, how to choose a probability to select the node is not trivial. In a classification scenario, the defender (i.e., the scheduling controller) attempts to maximize its accuracy, in contrast, the adversary wants to minimize the classifier’s accuracy, that is, maximize his attack success rate. This is a typical Stackelberg game and we can obtain a mixed strategy to determine the probability to select the node. Notice that besides the adversary, the user’s type as well includes legitimate users, so we further extend the Stackelberg game to the Bayesian Stackelberg Game as described in Section 2.3.

Now we formalize the Bayesian Stackelberg game in our EI-MTD as a seven-tuple $G_{Bayes}=\left(L, F^{(c)}, S_{L}, S_{F^{(c)}}, R_{L^{(c)}}, R_{F^{(c)}}, p^{(c)}\right)$, where $c \in \{1,2\}$. The defender $L$ is acted by the scheduling controller, and $S_{L}$ is his action set, i.e., choice of member models $\left\{F_{s}\left(\theta^{(k)}\right)\right\}_{k=1}^{K}$ for scheduling. Here, followers are end-users that including two types: legitimate users $F^{(1)}$ and adversaries $F^{(2)}$. $S_{F^{(1)}}$ is the legitimate user’s action set that has only one action that inputs a clean image for classification. Likewise, $S_{F^{(2)}}$ is the adversary’s action set that consists of multiple choices of substitute models $\left\{F_{a}\left(\theta^{(m)}\right)\right\}_{m=1}^{M}$ to generate adversarial examples. $R_{L^{(1)}}$ and $R_{F^{(1)}}$ are the payoff of the defender and legitimate users, which are defined as the classification accuracy of clean examples. $R_{L^{(2)}}$ is the payoff of defender under attacks, which is defined as the classification accuracy of adversarial examples. $R_{F^{(2)}}$ is the payoff of adversaries, which is defined as the attack success rate of adversarial examples. $P^{(1)}$ is the occurrence probability of a legitimate user and $P^{(2)}$ the occurrence probability of an adversary. The expected payoff functions of the legitimate user and the adversary are presented as follows respectively:

With the above definition, the model scheduling strategy is to maximize the defender’s payoff to obtain Bayes Nash equilibrium. As shown in Section 2.3, it is modeled as a MIQP problem: 
\begin{equation}
\begin{array}{l}
\max \limits_{s,q,v} \sum\limits_{n \in[K]}(P^{(1)} \cdot R_{L^{(1)}}(n, m)s_{n}q_{m}^{(1)} \\
\qquad\qquad+P^{(2)}\cdot\sum\limits_{m\in[M]} R_{L^{(2)}}(n, m) s_{n} q_{m}^{(2)})\\
\text{s.t.}\quad 0\leq v^{(c)}-\sum\limits_{n\in[K]}R_{F^{(c)}}(n,m) s_{n} \leq\left(1-q_{m}^{(c)}\right)N\\
\qquad\qquad\qquad\qquad\sum\limits_{n\in[K]}s_{n}=1\\
\qquad\qquad\qquad\qquad\sum\limits_{m\in[M]}q_{m}^{c}=1\\
\qquad\qquad\qquad\qquad c \in\{1,2\} \\
\qquad\qquad\qquad\qquad 0 \leq s_{n} \leq 1 \\
\qquad\qquad\qquad\qquad q_{m}^{(c)} \in\{0,1\} \\
\qquad\qquad\qquad\qquad v^{(c)} \in \mathbb{R}
\end{array}
\end{equation} 
where $P^{(1)}=1-\alpha$ and $P^{(1)}=\alpha$. Here $s$ is the solution of problem (11), which is a probability vector corresponding to member models for scheduling. For example, if $s=\left(p_{1}, p_{2}, \ldots, p_{K}\right)$ then $p_{i}$ is a probability corresponding to a member model $F_{s}(\theta^{(i)})$ to be chosen by the scheduling controller.

Since problem (13) is NP-hard, we use DOBSS [a26] to solve this problem. Compared to the other approaches, DOBSS has three key advantages. First, DOBSS allows for a Bayesian game to be expressed compactly without requiring conversion to a normal-form game via the Harsanyi transformation \cite{[a72]}. Second, DOBSS requires only one mixed-integer linear program to be solved, rather than a set of linear programs, thus the solution speed is further improved. Third, DOBSS directly searches for an optimal leader strategy, rather than a Nash equilibrium, thus allowing it to find high-payoff non-equilibrium strategies (exploiting the advantage of being the leader).

\begin{table}
\centering
\caption{The payoff matrix of the defender and legitimate users}
\begin{tabular}{lcc}
\multicolumn{3}{r}{Legitimate users}                                                                                         \\ \cline{2-3} 
\multicolumn{1}{c|}{}                          & \multicolumn{1}{c|}{MobileNetV2-1.0}  & \multicolumn{1}{c|}{(68.93, 68.93)} \\ \cline{2-3} 
\multicolumn{1}{l|}{}                          & \multicolumn{1}{c|}{MobileNetV2-0.75} & \multicolumn{1}{c|} {\textbf{(67.88, 67.88)}} \\ \cline{2-3} 
\multicolumn{1}{c|}{\multirow{2}{*}{Defender}} & \multicolumn{1}{c|}{ShuffleNetV2-0.5} & \multicolumn{1}{c|}{(64.98, 64.98)} \\ \cline{2-3} 
\multicolumn{1}{c|}{}                          & \multicolumn{1}{c|}{ShuffleNetV2-1.0} & \multicolumn{1}{c|}{(66.37, 66.37)} \\ \cline{2-3} 
\multicolumn{1}{l|}{}                          & \multicolumn{1}{c|}{SqueezeNet-1.0}   & \multicolumn{1}{c|}{(61.98, 61.98)} \\ \cline{2-3} 
\multicolumn{1}{l|}{}                          & \multicolumn{1}{c|}{SqueezeNet-1.1}   & \multicolumn{1}{c|}{(63.45, 63.45)} \\ \cline{2-3} 
\end{tabular}
\end{table}

\subsection{Theory Analysis}
If we deploy EI-MTD in real-world applications, we will determine all parameters including the occurrence probability of an adversary $\alpha$ , a distillation temperature $T$, and a regularization coefficient $\lambda$. Thus, we further utilize DOBSS to solve problem (11) and obtain a probability vector $s=\left(p_{1}, p_{2}, \ldots, p_{K}\right)$ for scheduling member models. Suppose the adversary’s target is model $F_{t}(\theta^{(k)})$, and he input adversarial examples $x_{adv}^{(i)}=x+r^{(i)},i=1,\ldots,M$ on edge device $e_{k}$, which are generated by $M$ substitute models. As described in Sec 3.1, the adversary records the success rate of all adversarial examples of the $i$-th substitute model as $c^{(i)}=\mathbb{I}\left(F_{t}\left(x_{a d v}^{(i)}, \theta^{(k)}\right) \neq y\right)$ and choose a proper substitute model $F_{a}\left(\theta^{(i^{*})}\right)=\arg \max _{i}\left\{\frac{1}{N} \sum_{n=1}^{N} c_{n}^{(i)}\right\}_{i=1}^{M}$ after $N$ queries. For the adversary, more queries can provide more accurate information statistically. Let $r(F_{a},F_{t})$ denote the transfer rate between models $F_{a}$ and $F_{t}$ where $F_{a}$ represent the substitute model to generate adversarial examples and $F_{t}$ is the target model.

\begin{theorem}
If a target model $F_{t}^{(k)}$ is determined with a probability $p_{k}$ in a dynamic scheduling setting, then the transfer rate $r_{\text{dymamic}}\left(F_{a}^{(i)}, F_{t}^{(k)}\right)$ in the dynamic setting does not converge to the transfer rate $r_{\text {static}}\left(F_{a}^{(i)}, F_{t}^{(k)}\right)$ in a static target model setting.
\end{theorem}

\begin{proof}
In a static target model setting, the adversary can obtain the expected transfer rate:
\begin{equation}
\begin{array}{l}
r_{\text {static}}\left(F_{a}^{(i)}, F_{t}^{(k)}\right)=\\
\mathbb{E}_{x_{adv, n}^{(i)}\sim P^{(i)}}\left[\frac{1}{N} \sum\limits_{n=1}^{N} \mathbb{I}\left(F_{t}\left(x_{a d v, n}^{(i)}, \theta^{(k)}\right) \neq y\right)\right]
\end{array}
\end{equation}
where $F_{a}^{(i)}$ is the $i$-th substitute model, $F_{t}^{(k)}$ is the target model, and $x_{adv,n}^{(i)}$ is an adversarial example at the $n$-th query and it subjects to a distribution $P^{(i)}$. Now we discuss the case with EI-MTD. Although the target model is still $F_{t}^{(k)}$ that the adversary wants to attacks, the transfer rate is changed since the true model for service is dynamically determined, and $F_{t}^{(k)}$ will be selected with the probability $p_{k}$. The adversary finally obtains the expected transfer rate:
\begin{equation}
\begin{array}{l}
r_{\text {dymamic}}\left(F_{a}^{(i)}, F_{t}^{(k)}\right)=\\
\mathbb{E}_{x_{adv, n}^{(i)} \sim P^{(i)}}\left[\frac{1}{N} \sum\limits_{k=1}^{K} \sum\limits_{n=1}^{N} \mathbb{I}\left(F_{t}\left(x_{adv, n}^{(i)}, \theta^{(k)}\right) \neq y\right) \times p_{k}\right]
\end{array}
\end{equation}
According to the definition of Bayesian Stackelberg equilibrium, the probability $p_{k}$ does not allow the adversary to achieve $r_{static}=r_{dymamic}$, so the adversary cannot find a proper substitute model.
\end{proof}

\section{EXPERIMENTS AND RESULTS}
In this section, we first describe the experiment setting and present comparative evaluation results of EI-MTD compared with other models against black-box attacks. We also analyze the impact of distillation temperature and the weight of different measures on EI-MTD.

\subsection{Experimental Setting}
In our experiments, we use a GPU cluster, a PC, and several Raspberry Pies to simulate a cloud data center, edge servers and edge devices in EI.

\begin{itemize}
    \item \textbf{Cloud data centers.} We use a GPU cluster that consists of 5 servers to simulate a cloud data center. Each server has four Geforce RTX 2080Tis and runs with Ubuntu 16.04.6LTS. We carry out adversarial training for teacher models and differential knowledge distillation for student models on this GPU cluster.
    \item \textbf{Edge servers.} We use a PC with Core i5-10210U 2.11GH and 16GB RAM to simulate an edge server. The scheduling controller runs on this edge server to schedule member models, which is implemented by python3.6 and PuLP2.1 on Windows 10.
    \item \textbf{Edge devices.} We use six Raspberry Pis with 64-bit four-core ARM Cortex-A53 and 1 GB LPDDR2 SDRAM, as edge devices. They connect to the edge server via Bluetooth. Besides student models on these edge devices, we develop a toy program to send images to the edge server at a random time point, which simulates image classification requests.
    \item \textbf{Datasets.} We conduct experiments on the ILSVRC2012 dataset \cite{[a75]}, which contains 1,000 categories and consists of 1.2 million images as a training set, and 150,000 images as the test set. The size of each image is 224$\times$224 with three color channels. It is currently a benchmark dataset in the field of image classification tasks.
    \item \textbf{Payoff Matrix.} A payoff matrix in a game represents the payoff of participants under different strategy profiles. The element of the payoff matrix is a two-tuple $(a,b)$, where $a$ is the accuracy of the classifier under attacks while $b$ is the attack success rate. We obtain $a$ and $b$ through testing the member models and substitute models with the test set of ILSVRC2012. However, for legitimate users, their payoffs are the accuracy of the classifier since their goal is not to attack the classifier. Table 1 shows the game payoffs between the defender and legitimate users in the MTD-EI framework. Tables 2 presents the payoff matrix between the defender and the adversaries (PGD). For example, (56.73, 43.27) in Table 2 means that the defender MobilenetV2-1.0 achieves 56.73\% accuracy as its payoff under PGD adversarial attacks and the adversary achieves 43.27\% attack success rate as his payoff with PGD adversarial examples generated on the substitute model ResNet-18. Due to the limitation of paper size, we do not present other payoff matrices of adversaries such as FGSM, MI-FGSM and M-DI$^{2}$-FGSM. 

\end{itemize}

\begin{table*}
\centering
\caption{The payoff matrix of the defender and the PGD adversary}
\resizebox{\textwidth}{!}{
\begin{tabular}{|c|c|c|c|c|c|c|}
\hline
\multicolumn{2}{|c|}{\multirow{2}{*}{}}       & \multicolumn{5}{c|}{PGD adversary}                                                                                                                                                         \\ \cline{3-7} 
\multicolumn{2}{|c|}{}                        & \multicolumn{1}{c|}{MobileNetV2-1.0} & \multicolumn{1}{c|}{ShuffleNetV2-1.0}        & \multicolumn{1}{c|}{ResNet-18}               & \multicolumn{1}{c|}{SuqeezeNe-1.0}  & VGG-13         \\ \hline
\multirow{6}{*}{Defender} & MobileNetV2-1.0  & \multicolumn{1}{c|}{(25.92, 74.08)}  & \multicolumn{1}{c|}{(53.43, 46.57)}          & \multicolumn{1}{c|}{\textbf{(56.73, 43.27)}} & \multicolumn{1}{c|}{(51.66, 48.34)} & (50.53, 49.47) \\ \cline{2-7} 
                          & MobileNetV2-0.75 & \multicolumn{1}{c|}{(30.06, 69.94)}  & \multicolumn{1}{c|}{(50.74, 49.26)}          & \multicolumn{1}{c|}{(54.57, 45.43)}          & \multicolumn{1}{c|}{(48.72, 51.28)} & (49.86, 50.14) \\ \cline{2-7} 
                          & ShuffleNetV2-0.5 & \multicolumn{1}{c|}{(52.77, 47.23)}  & \multicolumn{1}{c|}{\textbf{(27.16, 72.84)}} & \multicolumn{1}{c|}{(51.75, 48.25)}          & \multicolumn{1}{c|}{(50.62, 49.48)} & (57.14, 42.86) \\ \cline{2-7} 
                          & ShuffleNetV2-1.0 & \multicolumn{1}{c|}{(50.22, 49.78)}  & \multicolumn{1}{c|}{\textbf{(18.62, 81.38)}} & \multicolumn{1}{c|}{(49.33, 50.67)}          & \multicolumn{1}{c|}{(54.02, 45.98)} & (53.77, 46.23) \\ \cline{2-7} 
                          & SqueezeNet-1.0   & \multicolumn{1}{c|}{(40.54, 59.46)}  & \multicolumn{1}{c|}{(42.19, 57.81)}          & \multicolumn{1}{c|}{(41.42, 58.58)}          & \multicolumn{1}{c|}{(28.81, 71.19)} & (44.14, 55.86) \\ \cline{2-7} 
                          & SqueezeNet-1.1   & \multicolumn{1}{c|}{(41.31, 58.69)}  & \multicolumn{1}{c|}{(45.20, 54.80)}          & \multicolumn{1}{c|}{(38.97, 61.03)}          & \multicolumn{1}{c|}{(31,13, 68.87)} & (40.06, 59.94) \\ \hline
\end{tabular}}
\end{table*}

\subsection{Results of Adversarial Training and Differential Distillation}
To ensure a higher degree of knowledge transfer from a teacher model to student models, the teacher model itself must have enough accuracy and robustness. We first evaluate the accuracy of the teacher model with adversarial training and then evaluate the accuracy of the student models.

\textbf{Accuracy of the teacher model with adversarial training.} We adopt ResNet-101 \cite{[a29]} as the teacher model in our experiment. ResNet-101 has 101 layers and 33 residual blocks. It is trained on our GPU cluster with 1.2 million clean images and their corresponding adversarial examples crafted by FGSM as processed in \cite{[a15]}. We use 150,000 clean images and 10,000 adversarial images to test the accuracy of the teacher model at each training epoch. Notice that the training data and test data are independent, and the adversarial examples for training are crafted by FGSM with $\varepsilon = 10$, while the test adversarial images are crafted by PGD with the perturbation bound $\varepsilon = 5$, the iteration step size is $\varepsilon / 5$, and the number of iterations is 20. We use $\varepsilon = 0.03$, 10 PGD steps for training and 40 steps with random restart for evaluation. Fig. 3 demonstrates the accuracy during the adversarial training. With the increase of epochs, the accuracy of the teacher model is gradually improved, which reconfirm the effectiveness of adversarial training on the cloud center. After total 15 epochs, our teacher model achieved both $\thicksim$83\% clean top-5 accuracy and $\thicksim$74\% adversarial top-1 accuracy, which can ensure our differential knowledge distillation to achieve perfect effects.

\begin{figure}
	\centering  
	\includegraphics[width=6cm,height=4.3cm]{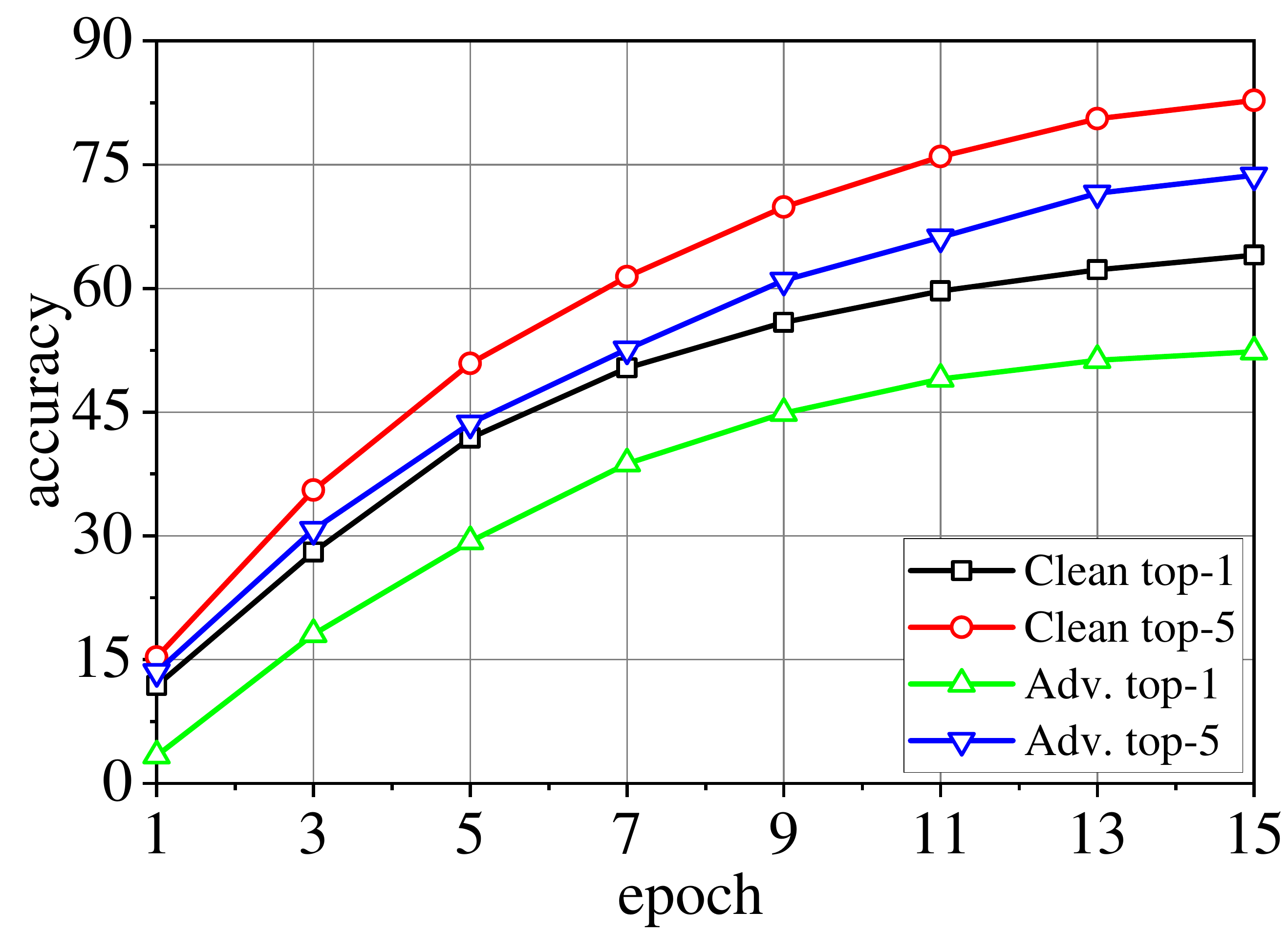}  
	\caption{The top-1 and top-5 accuracy of the teacher model during adversarial training. After each training epoch, the teacher model is tested with two data sets that the one includes clean images and the other includes PGD adversarial examples. }  
	\label{fig:mcmthesis-logo}   
\end{figure}

\begin{figure} 
\subfigure[Clean Top-1]{
\includegraphics[width=3.9cm,height=2.8cm]{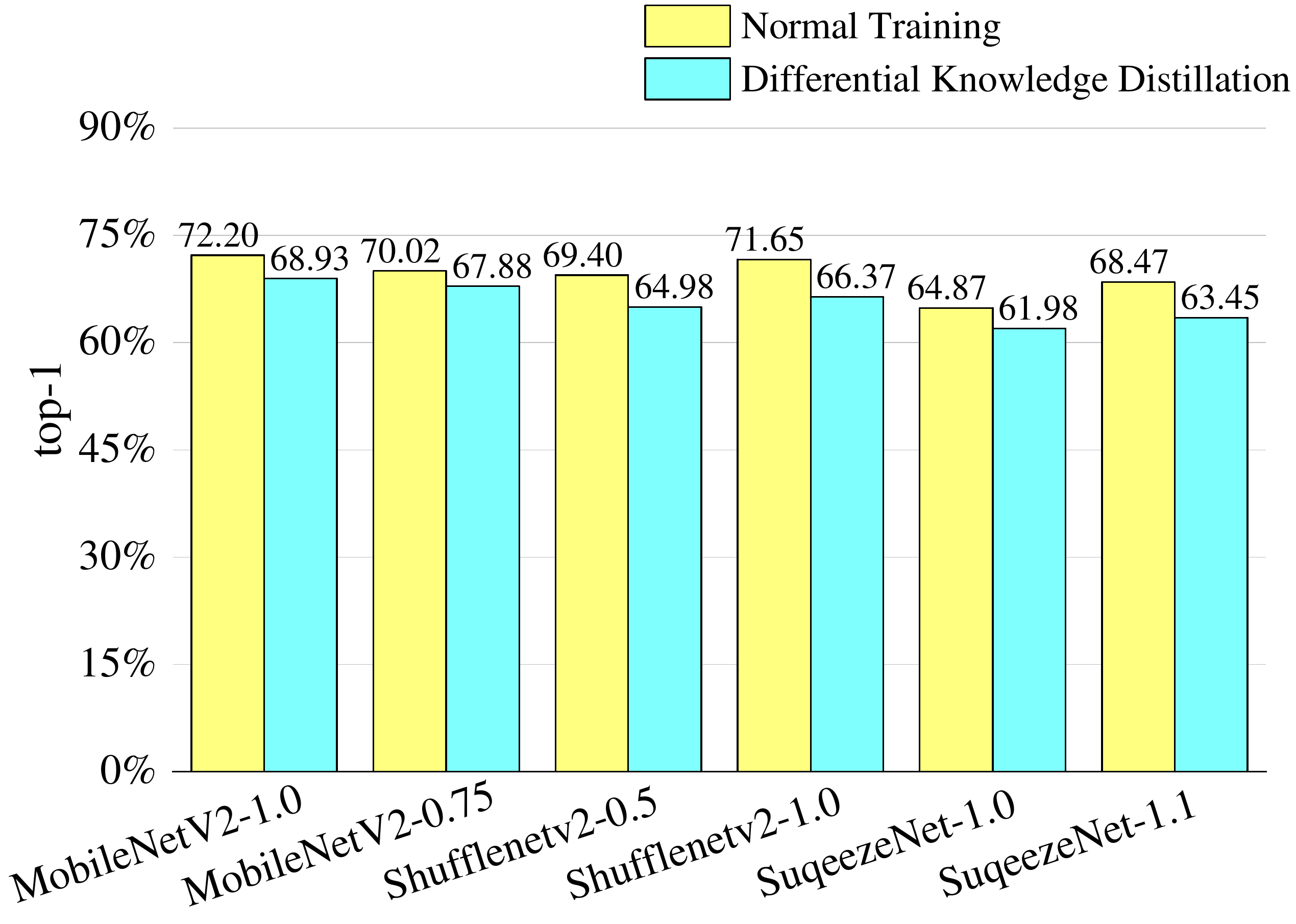} 
}
 \subfigure[Clean Top-5]{
\includegraphics[width=3.9cm,height=2.8cm]{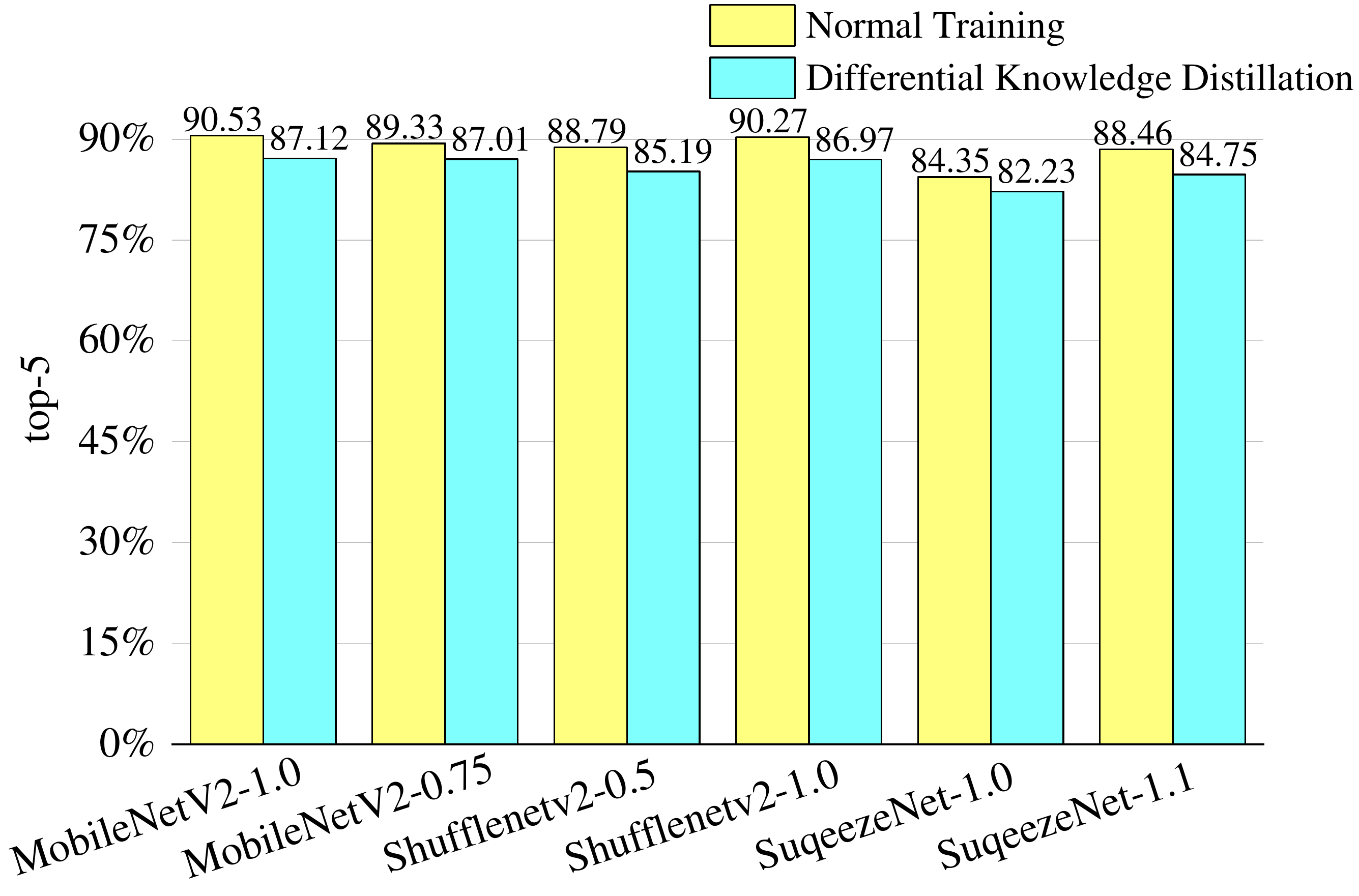} 
} 

 \subfigure[Adv. Top-1]{
\includegraphics[width=3.9cm,height=2.8cm]{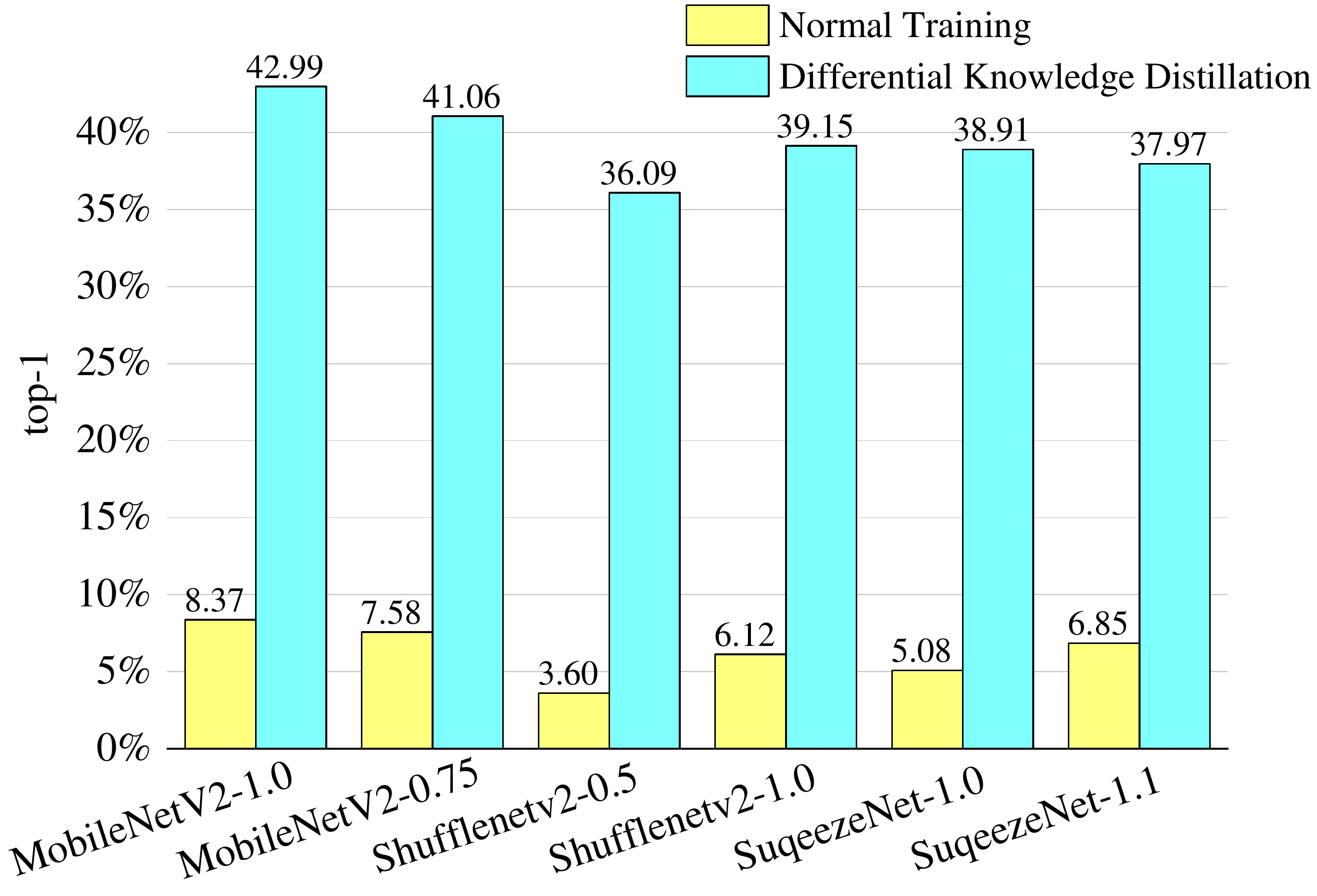}
 }
 \subfigure[Adv. Top-5]{
\includegraphics[width=3.9cm,height=2.8cm]{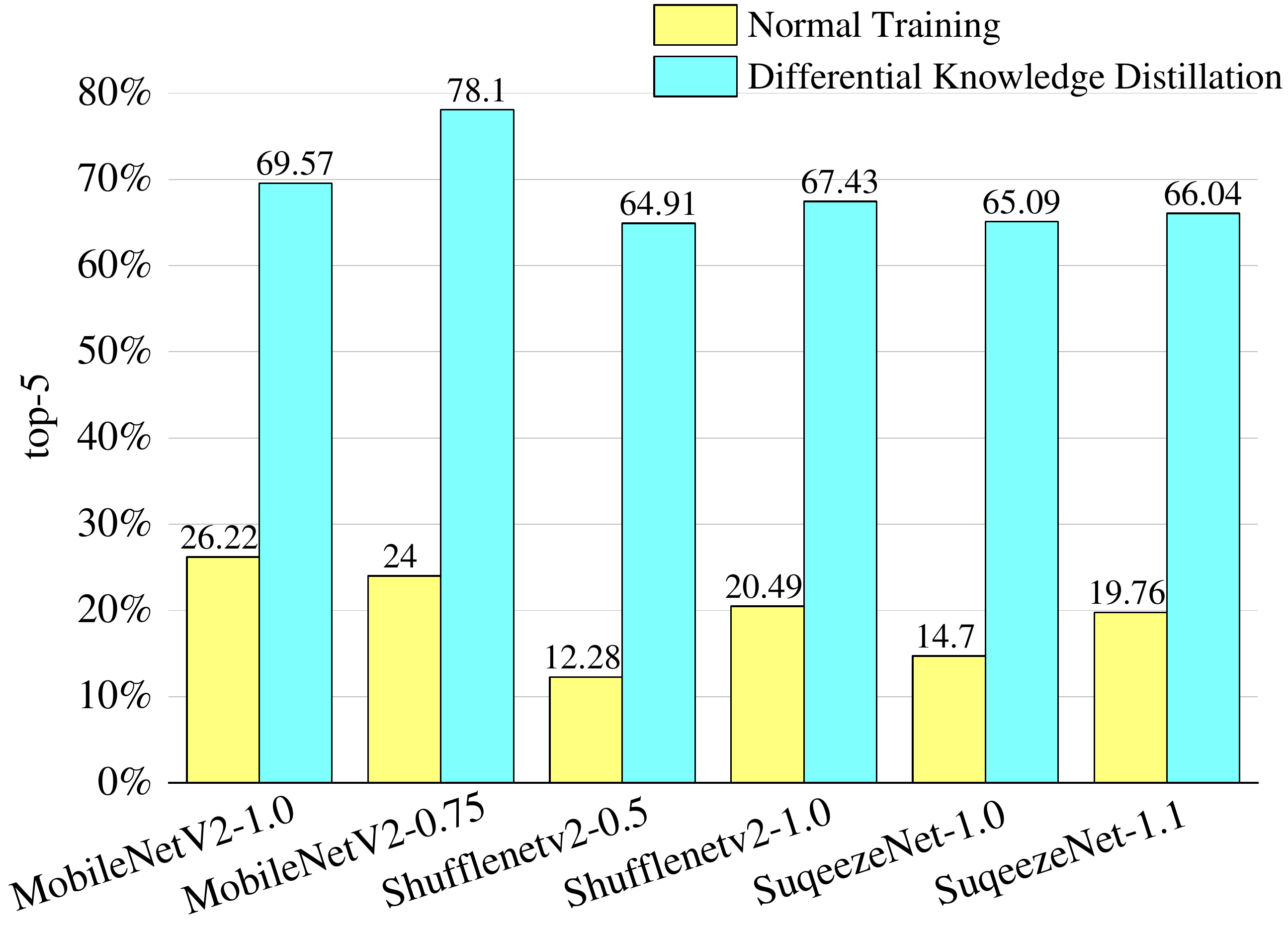}
 }
 \caption{Accuracy of member models between normal training and differential knowledge distillation}
 \end{figure}

\textbf{Accuracy of the student models with differential distillation.} We adopt several current mainstream lightweight model structures, namely MobileNetV2 \cite{[a76]}, ShuffleNetV2 \cite{[a77]} and SqueezeNet \cite{[a78]} for student models. To obtain more student models, we further tune the super-parameters of these models and the finally student models are MobileNetV2-1.0, MobileNetV2-0.75, ShuffleNetV2-0.5, ShuffleNetV2-1.0, SqueezeNet-1.0, and SqueezeNet-1.1. For comparison, two group models are trained. The first group is by normal training and the other group is by differential knowledge distillation. We use 150,000 clean images and 10,000 PGD adversarial images to test these two group student models. The results are shown in Fig. 4. We observe that the accuracy of the models with normal training is little lower than that of the models with differential knowledge distillation on clean images. However, the models with differential knowledge distillation are far more robust than the models with normal training by $\thicksim$33\% top-1 accuracy and $\thicksim$49\% top-5 accuracy in average on adversarial images. These results indicate that the student models distilled from a robust teacher model have preferable defensive performance against adversarial examples regardless of their reduced size.

\subsection{Effectiveness of EI-MTD}
To simulate the adversary’s black-box attack strategy, we assume the adversary has five substitute models: MobileNetV2-1.0, Shufflenetv2-1.0, SuqeezeNet-1.0, ResNet-18\cite{[a29]}, and VGG-13\cite{[a79]}. Among them, the first three models are similar to the member models for simulating the static scenario in which the adversary can find a similar substitute model. Thus, we generate FGSM, PGD, MI-FGSM and M-DI$^{2}$-FGSM adversarial examples on these substitute models to test our proposed EI-MTD. Specifically, the perturbation bound $\varepsilon=10$ for FGSM, the step size is $\varepsilon/10$ for MI-FGSM, $\varepsilon/5$for PGD and the decay factor $\mu=0.5$ for M-DI$^{2}$-FGSM, and the perturbation is randomly initialized with $U(-\varepsilon,+\varepsilon)$.In addition, for M-DI$^{2}$-FGSM, the stochastic transformation function $T(x,\rho)$ is used to resize the input $x$ randomly to a $rand \times rand \times 3$image, with $\rho=0.5$ and $rand \in [299,300)$.

\textbf{Accuracy of EI-MTD under black-box attacks.} Given the payoff matrixes, we can obtain the defender’s optimal mixed strategy through solving Eq. (11), i.e., a probability vector for selecting a proper member model. Since the defender’s optimal mixed strategy depends on the occurrence probability $\alpha$ of an adversary, we discussed the effectiveness of EI-MTD according $\alpha$. In essentially, $\alpha$ is the ratio of adversarial examples in the total test examples. For comparison, we test 5 member models independently to simulate a traditional static target model scheme as shown in the left of Fig. 1. Fig. 5 demonstrates the results as follows: 

\begin{itemize}
  \item [(1)] \pmb{$\alpha = 0$}. This case means that the user type has only legitimate users, which is essentially equivalent to a normal classification with clean examples. Here we use 10,000 clean images for test. As a result, the scheduling controller uses a pure strategy to select the member model with the highest accuracy of clean examples. For instance, the member model Mobilenetv2-1.0 is selected for service as shown in Fig 5(b).

  \item [(2)] \pmb{$\alpha = 1$}. This case assumes that the user type has only adversaries, which means that all classification requests are adversarial examples. Here we use 10,000 adversarial images for test. In our experiment, the scheduling controller obtains a probability vector $s=(0.12, 0.15, 0.19, 0.09, 0.14, 0.31)$ by solving Eq. (10) with $\alpha=1$. The probabilities of this vector are respectively corresponding to 6 member models as presented in Table 3. Then the controller selects a model according to its corresponding probability. As shown in Fig 5, even if all the examples are adversarial, the expected accuracy of EI-MTD is still higher than that of any static target models.

  \item [(3)] \pmb{$0<\alpha<1$}. In a practical scenario, the legitimate user and the adversary exist with a certain ratio denoted by $\alpha$. In our experiment, we set $\alpha=0.1,0.2,\ldots,0.9$ to simulate various possible scenarios, e.g., $\alpha=0.1$ means 1,000 adversarial images and 9,000 clean images in our total 10,000 test images. Similar to $\alpha=1$, given a determined $\alpha$, the controller will compute a probability vector for selecting a target model. The results with various $\alpha$ are shown in Fig. 5. As $\alpha$ increases, i.e., the proportion of adversarial examples in the request is improved, the accuracy of all models including EI-MTD decrease almost linearly. It is not surprised to us since increasing adversarial examples is able to increase the attack success rate when the number of test examples remains unchanged. No matter how $\alpha$ changes, the classification accuracy of EI-MTD is $\thicksim$15\% higher than that of any single target model.

 \end{itemize}
 
As analyzed above, we can summarize that no matter what adversarial examples (FGSM, PGD, MI-FGSM, and M-DI$^{2}$-FGSM) are used for test and no matter what proportion the adversarial examples are, EI-MTD outperforms a static target model.

\renewcommand\arraystretch{1.5}   
\begin{table}
\centering
\caption{The probability for choice of member models when $\alpha=1$}
\setlength{\tabcolsep}{0.2mm}   
\begin{tabular}{|c|c|c|c|}
\hline
Member Models & MobileNetV2-1.0  & MobileNetV2-0.75 & ShuffleNetV2-0.5 \\ \hline
Probability   & 0.12             & 0.15             & 0.19             \\ \hline
\hline
Member Models & ShuffleNetV2-1.0 & SqueezeNet-1.0   & SqueezeNet-1.1   \\ \hline
Probability   & 0.09             & 0.14             & 0.31             \\ \hline
\end{tabular}
\end{table}

\begin{figure*} 
\centering 
\subfigure[FGSM]{
\includegraphics[width=4cm,height=2.8cm]{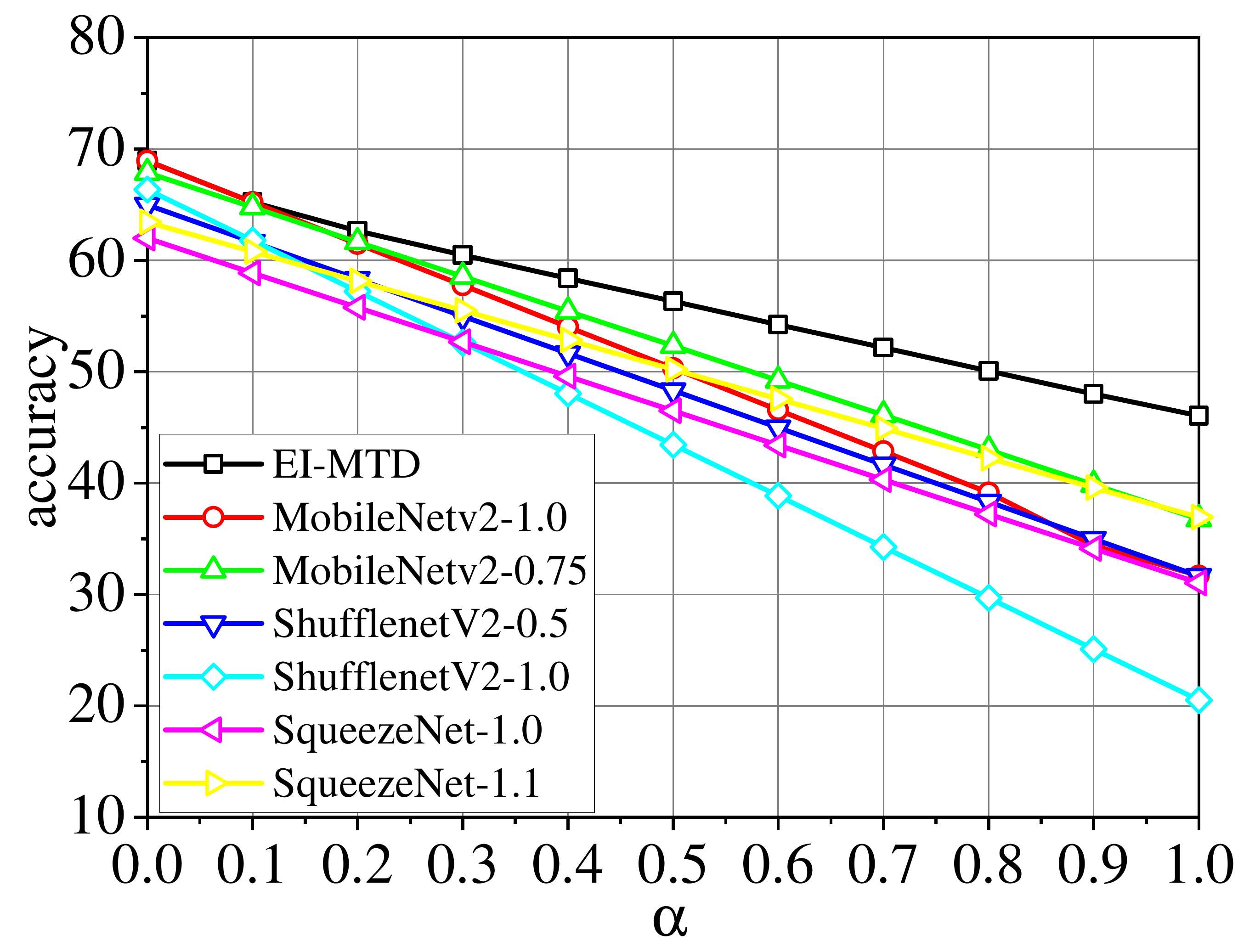} 
}
 \quad
 \subfigure[PGD]{
\includegraphics[width=4cm,height=2.8cm]{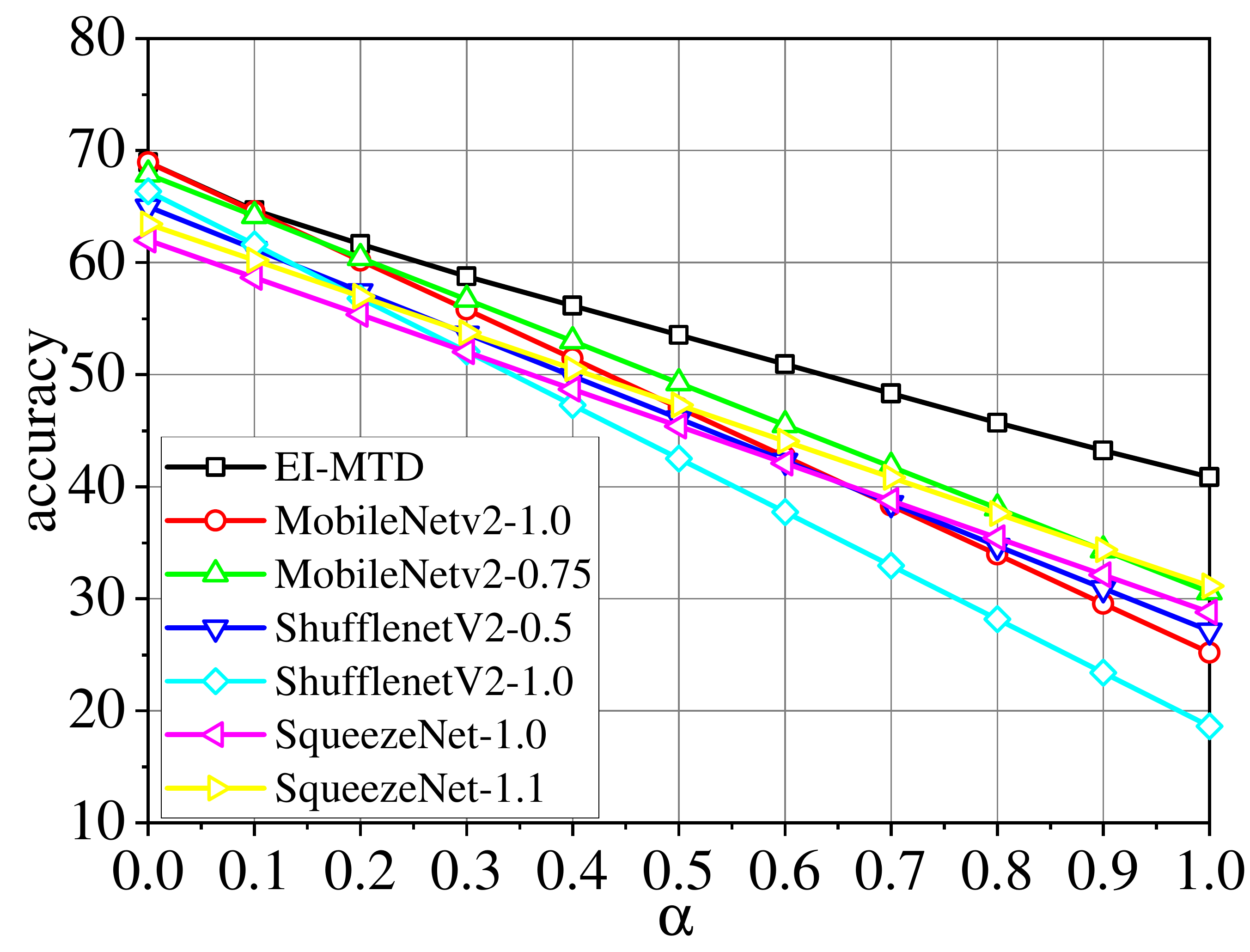} 
} 
 \quad
 \subfigure[MI-FGSM]{
\includegraphics[width=4cm,height=2.8cm]{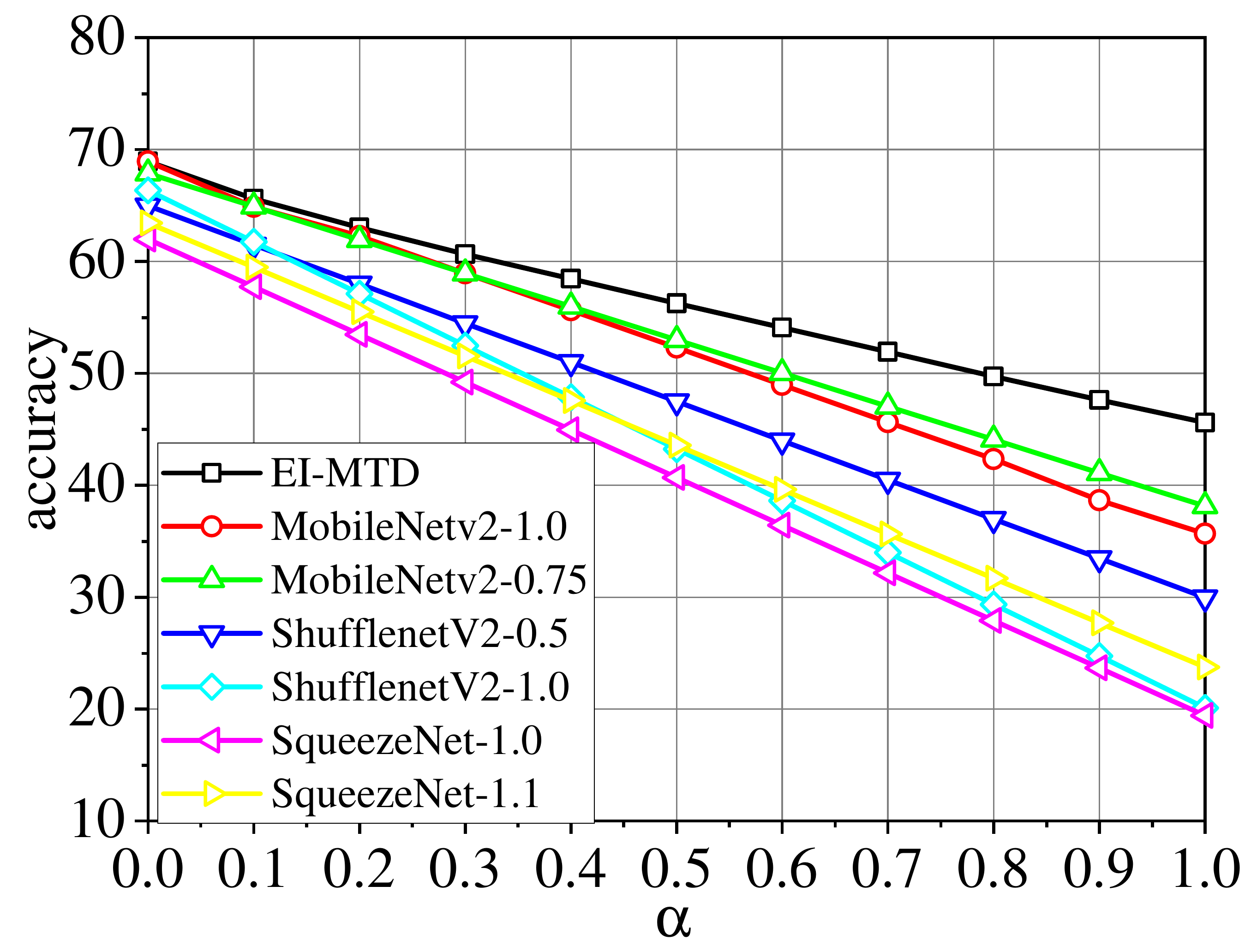}
 }
  \quad
 \subfigure[M-DI$^{2}$-FGSM]{
\includegraphics[width=4cm,height=2.8cm]{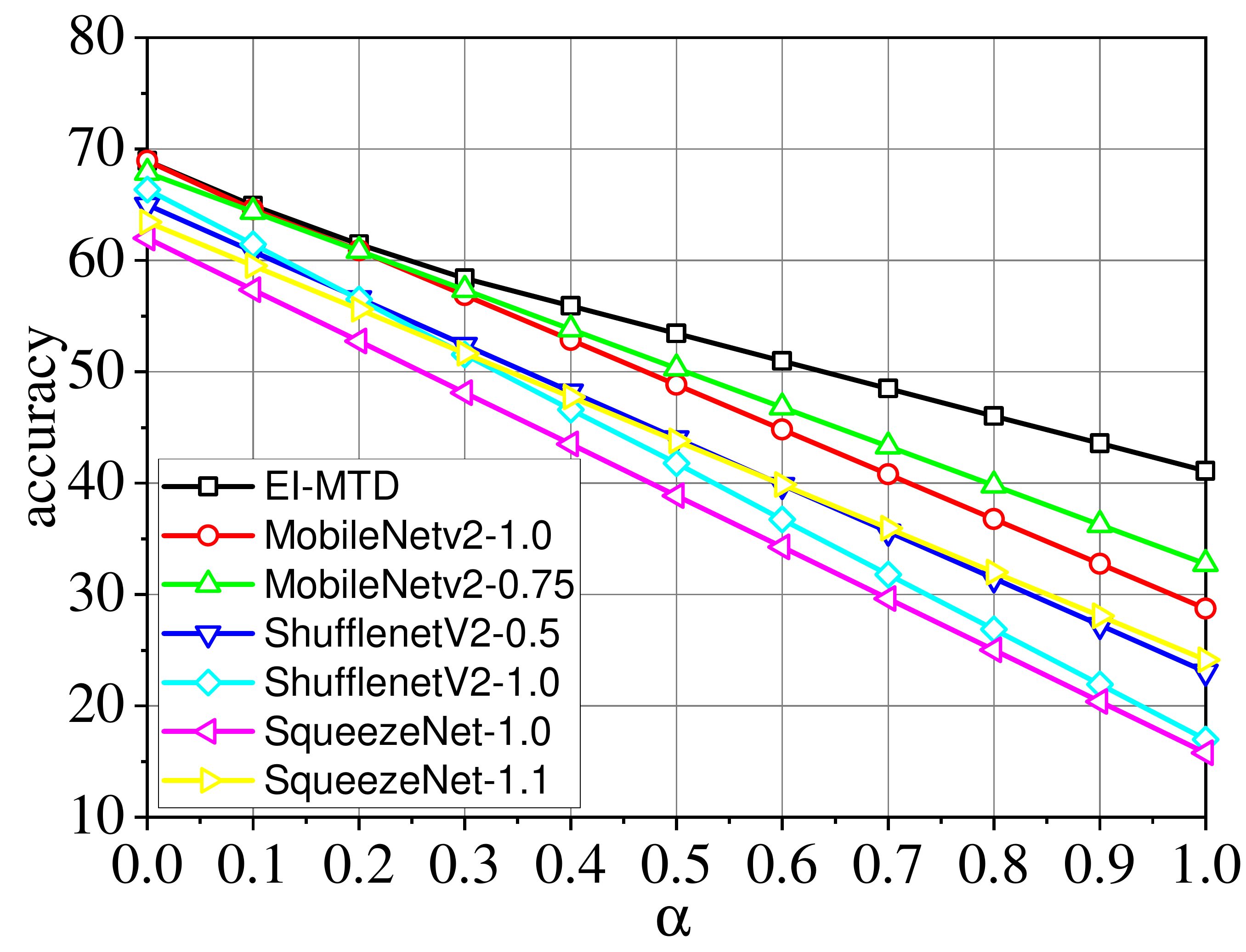}
 }
 \caption{The horizontal axis represents the ratio of adversarial examples. The vertical axis is the accuracy of EI-MTD and static target models for comparison. (a), (b), (c) and (d) are respectively corresponding to the adversarial examples generated by FGSM, PGD, MI-FGSM, and M-DI$^{2}$-FGSM for test. Our proposed EI-MTD outperforms any of static target models under these typical adversarial attacks.}
 \end{figure*}

\textbf{Transferability on EI-MTD.} The transferability of adversarial examples can be measured by the transfer rate, i.e., the ratio of the number of transferred adversarial examples to the total number of adversarial examples constructed by the original model \cite{[a65]}. In our experiment, the percentage of adversarial examples correctly classiﬁed by our EI-MTD or a single member model generated is used to measure the transferability. Notice these adversarial examples are generated by the adversary’s substitute models. Essentially, the transfer rate is equivalent to the value of 100\% subtracting the accuracy of the target model. Thus, we use the accuracy in Fig. 5 with respect to $\alpha$ that represents all test examples are adversarial examples to compute the transfer rate. Obviously, the transfer rate of adversarial examples on EI-MTD is lower than that on other member models. For instance, in Fig. 5d, the transfer rate on EI-MTD is 58.91\%, while that on the member model such as MobileNetV2-1.0 is 71.26\%. Similarly, the transfer rate on all other member models is higher than EI-MTD, which indicates the efficiency of EI-MTD against transferability.

\subsection{Impact of $T$ And $\lambda$ on EI-MTD}
To gain insight into the impaction of differential distillation on EI-MTD, we further analyze two important parameters $T$ and $\lambda$, which respectively represent a distillation temperature and a regularization coefficient in Eq. (10). Sailik et al. \cite{[a23]} propose differential immunity as a measure for the effectiveness of MTD based on the point that for an ideal MTD, a specific attack only effective on one particular conﬁguration, but ineffective for all the others. Thus, they define differential immunity with respect to the attack success rate (ASR) as follows:
\begin{equation}
\gamma=\frac{\max _{F_{s}} ASR\left(F_{a}, F_{s}\right)-\min _{F_{s}} ASR\left(F_{a}, F_{s}\right)+1}{\max _{F_{s}} ASR\left(F_{a},F_{s}\right)+1}
\end{equation}
where $F_{a} \in U$ denotes a substitute model, $F_{s} \in \Omega$ denotes a target model selected by the scheduling controller, and $ASR(F_{a},F_{s})$ represents the attack success rate of adversarial examples generated from model $F_{a}$ against the model $F_{s}$. Consequently, a larger $\gamma$ indicates a fine performance of MTD. In this section, we use differential immunity to explore the impact of $T$ and $\lambda$ on EI-MTD.



\begin{figure*}
\begin{minipage}[t]{0.306\textwidth}
	\centering  
	\includegraphics[width=5cm,height=3.8cm]{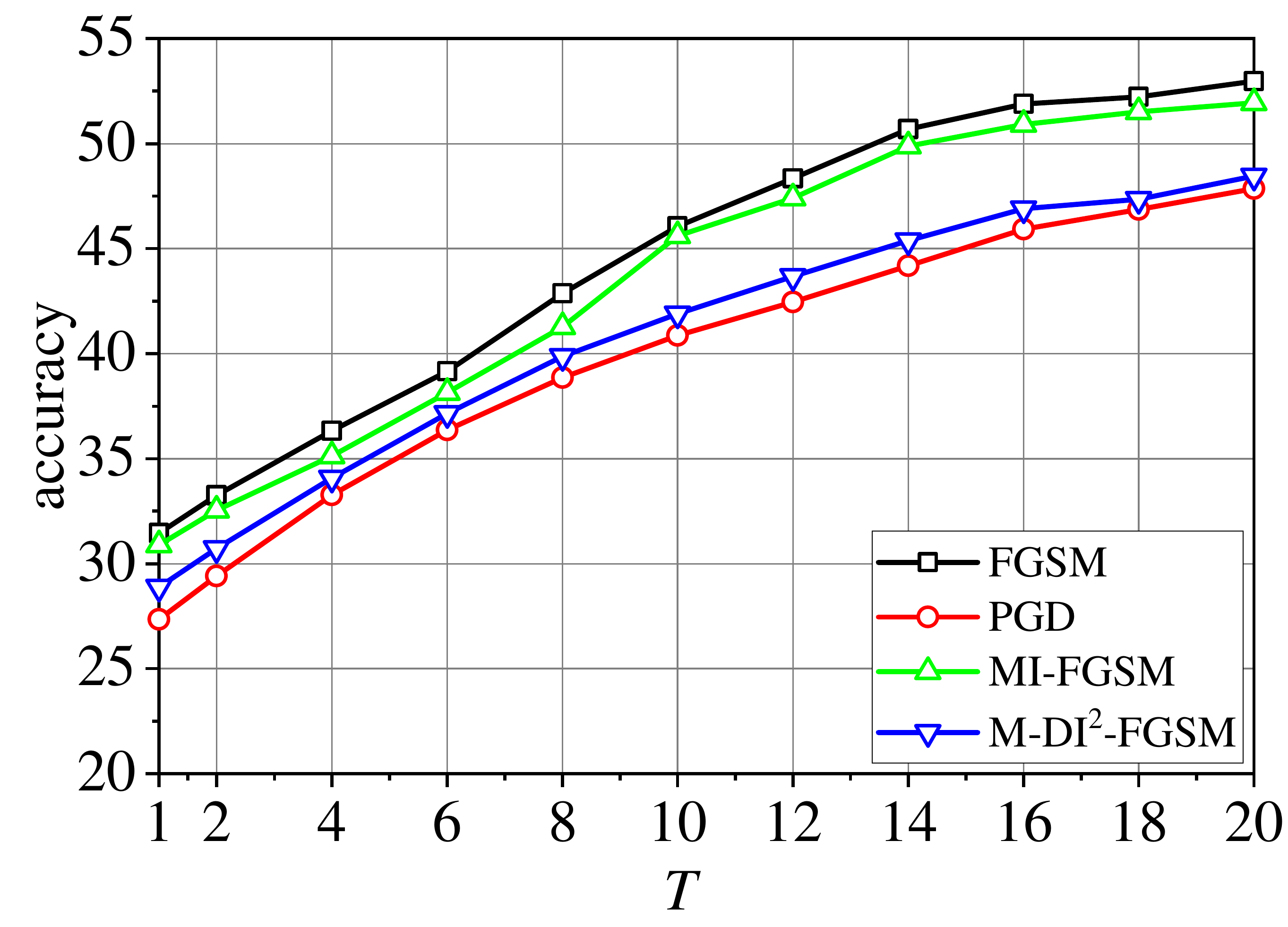}  
	\caption{The accuracy of EI-MTD with respect to distillation temperature $T$ ($\lambda$=0.3, $\alpha$=1)}  
	\label{fig6}   
\end{minipage}
\qquad
	\begin{minipage}[t]{0.306\textwidth}
	\centering  
	\includegraphics[width=5cm,height=3.8cm]{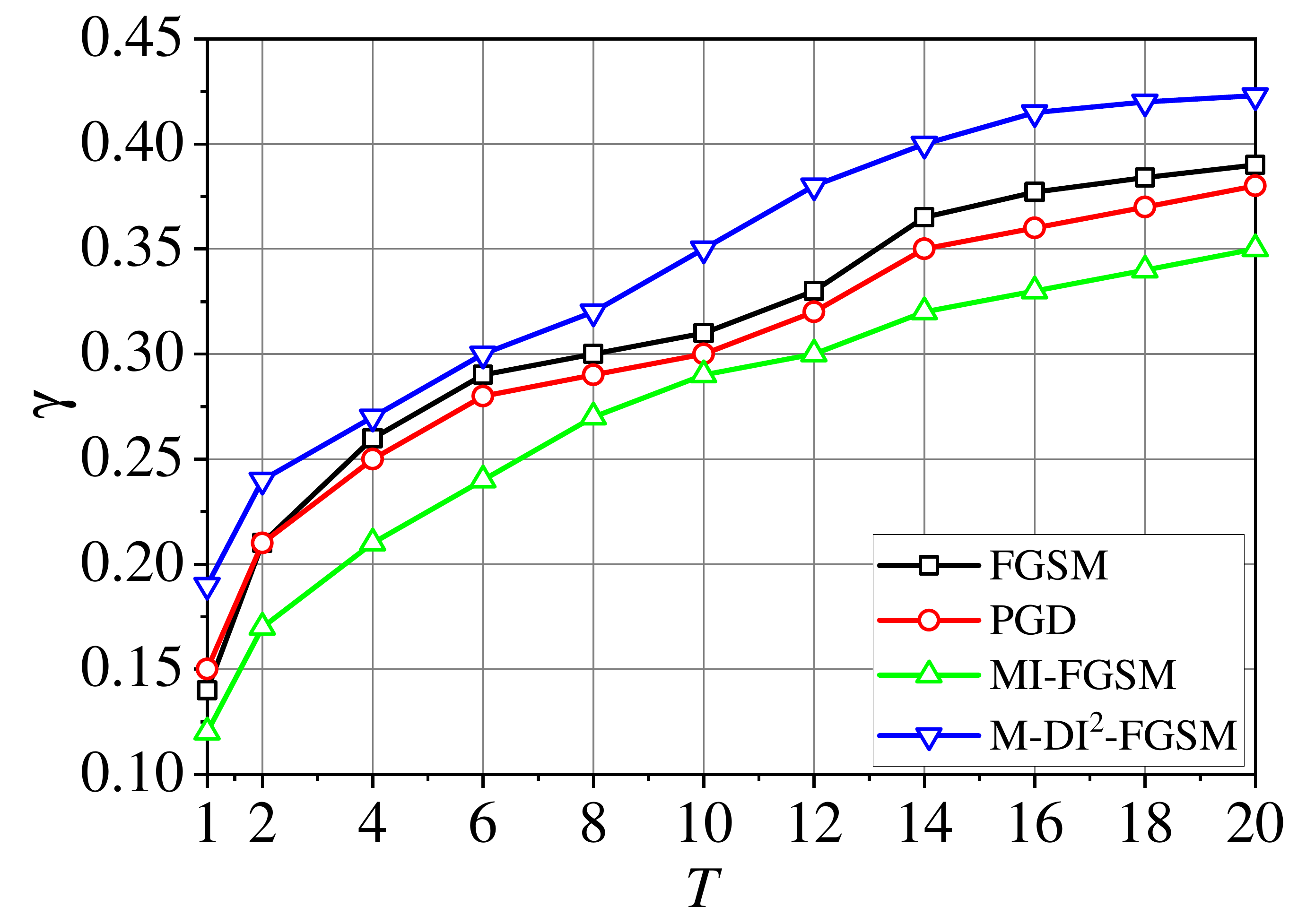}  
	\caption{The differential immunity $\gamma$ with respect to the distillation temperature $T$ ($\lambda$=0.3, $\alpha$=1)}  
	\label{fig7}   
	\end{minipage}
\qquad
\begin{minipage}[t]{0.306\textwidth}
	\centering  
	\includegraphics[width=5cm,height=3.8cm]{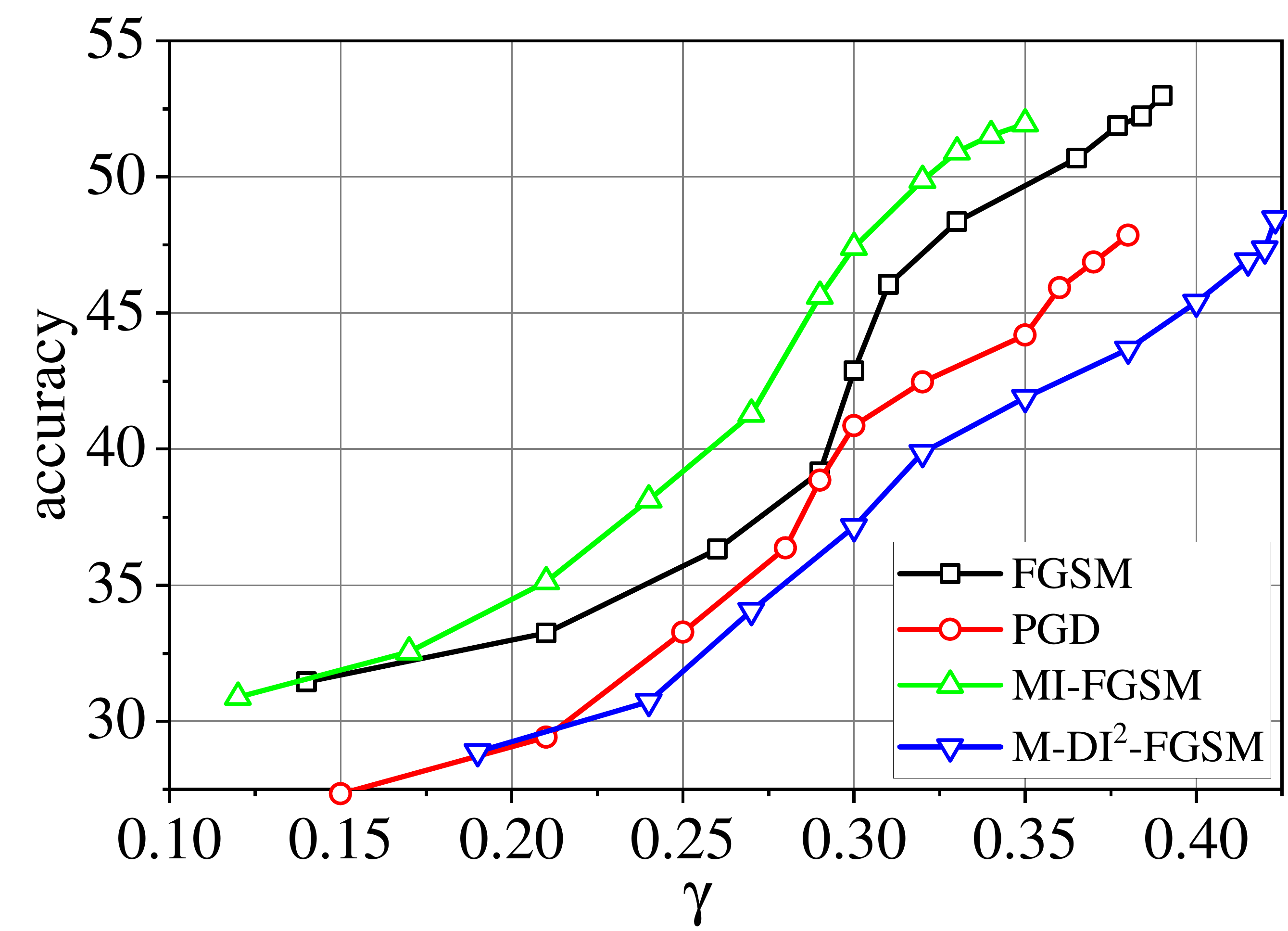}  
	\caption{The accuracy of EI-MTD with respect to the differential immunity $\gamma$ ($\lambda$=0.3, $\alpha$=1)}  
	\label{fig8}   
\end{minipage}
\end{figure*}

\begin{figure*}
\begin{minipage}[t]{0.306\textwidth}
	\centering  
	\includegraphics[width=5cm,height=3.8cm]{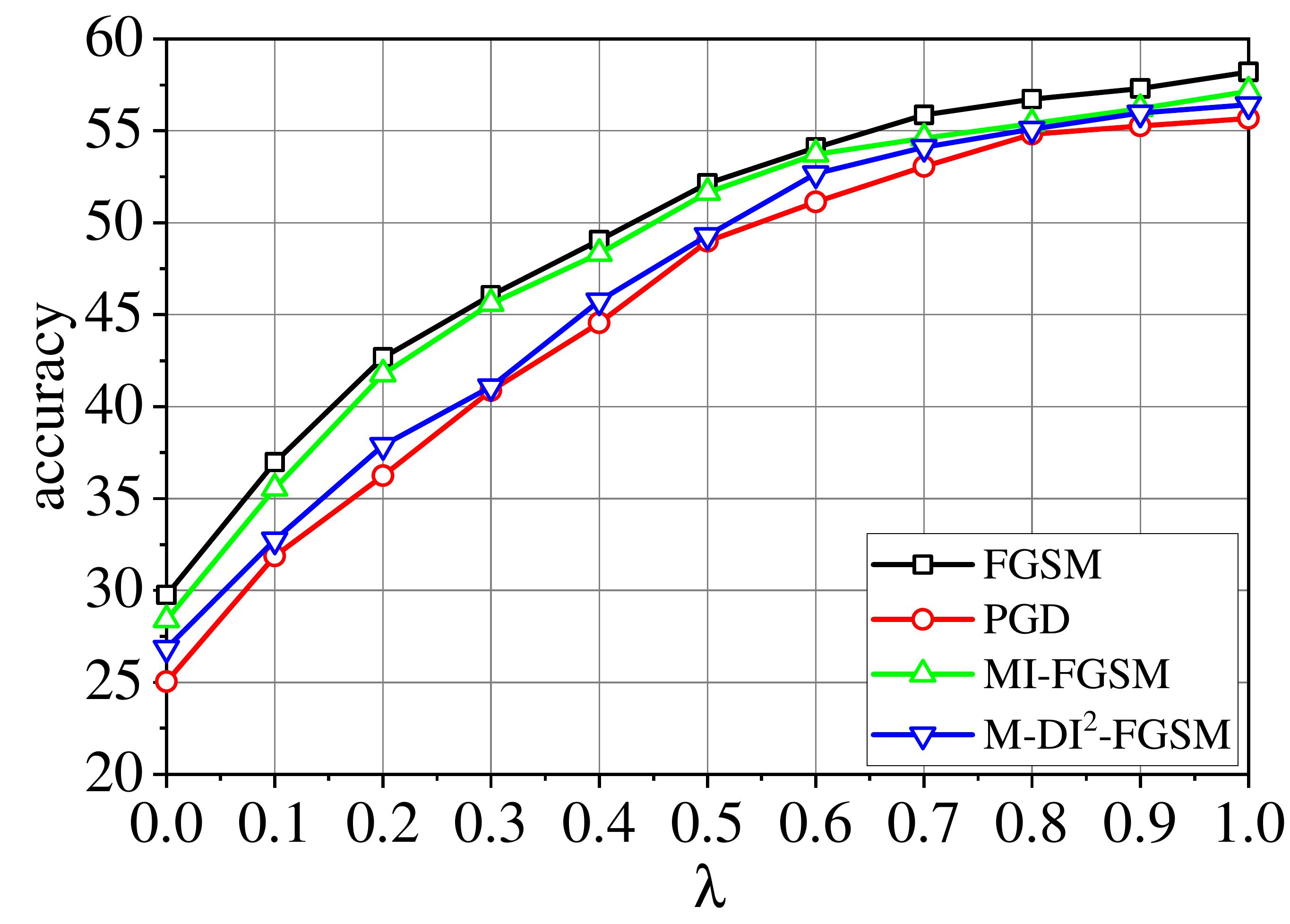}  
	\caption{The accuracy of EI-MTD with respect to the regularization coefficient $\lambda$ ($T$=10, $\alpha$=1)}  
	\label{fig9}   
	\end{minipage}
\qquad
\begin{minipage}[t]{0.306\textwidth}
	\centering  
	\includegraphics[width=5cm,height=3.8cm]{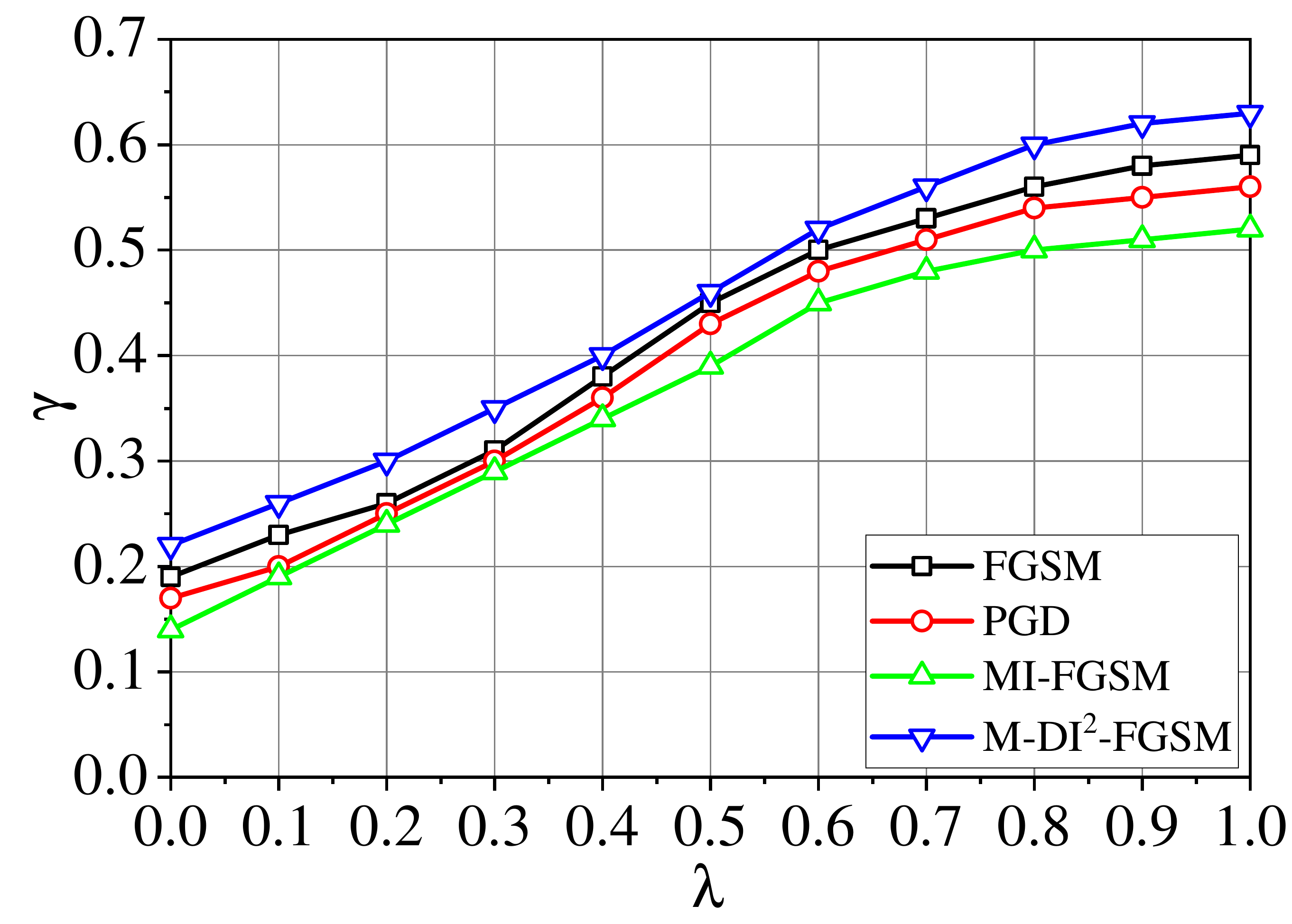}  
	\caption{The differential immunity $\gamma$ with respect to regularization coefficient $\lambda$ ($T$=10)}  
	\label{fig10}   
\end{minipage}
\qquad
\begin{minipage}[t]{0.306\textwidth}
	\centering  
	\includegraphics[width=5cm,height=3.8cm]{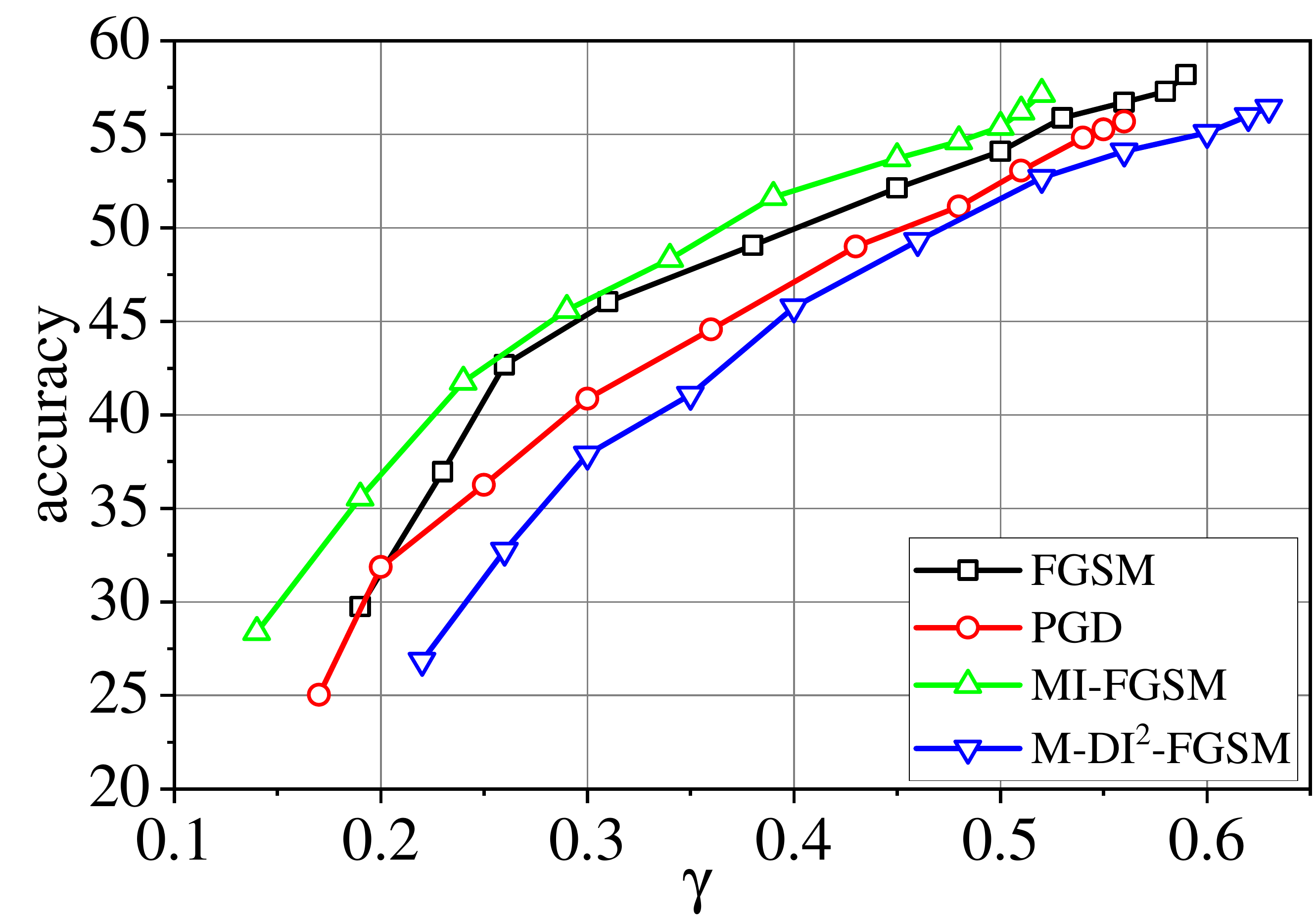}  
	\caption{The accuracy of EI-MTD with respect to the differential immunity $\gamma$ ($T$=10)}  
	\label{fig11}   
\end{minipage}
\end{figure*}

\textbf{The impact of $T$.} For conveniently investigating $T$, we first fix $\lambda=0.3$. We also assume that all the requests are adversarial examples and let $\alpha=1$. The relationship between the accuracy of EI-MTD and distillation temperatures was shown in Fig. \ref{fig6}. With the increase of distillation temperature, we observed the accuracy of EI-MTD is improved accordingly regardless of any adversarial examples including FGSM, PGD, MI-FGSM and M-DI$^{2}$-FGSM. This observation further inspires us to explore the intrinsic link between distillation temperatures and the accuracy of EI-MTD.

Since we have obtained the accuracy of all member models, the differential immunity can be easily computed by Eq. (14). Fig. \ref{fig7} presents the differential immunity with respect to the distillation temperature. We observed that the differential immunity $\gamma$ increased with the distillation temperature rising, which implies that higher distillation temperature can enlarge the diversity of member models. The reason is that higher $T$ can guide the decision boundary of member models close to the robust teacher model, which reduces the denominator in Eq. (7). However, after $T$ increasing to 12, the growth of $\gamma$ becomes flat, it indicates the distillation temperature may be no longer the main factor of member model differences at this time. Hence it is favorable evidence to explain the effectiveness of our EI-MTD.

Based on the above observation, we further analyze the link between the accuracy of EI-MTD and differential immunity $\gamma$. In Fig. \ref{fig8}, we present the accuracy with respect to different $\gamma$ by experiments. It shows that increasing $\gamma$ can improve the accuracy of EI-MTD, which reconfirms our point described in Section 4.1 that the diversity of member models determines the effectiveness of EI-MTD. For example, when $\gamma=0.15$, the accuracy of EI-MTD is only 27.34\%, but when we increasing $\gamma$ to 0.38 by differential knowledge distillation with $T=20$, EI-MTD achieves 47.86\% accuracy. Accordingly, we can clearly explain how the distillation temperature $T$ impact on the effectiveness of EI-MTD: (1) increasing $T$ can increase the differential immunity $\gamma$; (2) increasing $\gamma$ can further improve the accuracy; so (3) increasing $T$ finally can improve the effectiveness of EI-MTD.

\textbf{The impact of $\lambda$.} Remind $\lambda$ is a regularization coefficient that controls the importance given to $CS_{coherence}$ during training. To reveal the impact of $\lambda$ on EI-MTD performance, we fixed the distillation temperature $T=10$. As shown in Fig. 9, increasing $\lambda$ can perfectly improve the accuracy of EI-MTD, which is exactly what we expected. However, this result is not direct to reveal their relationship essentially. Thus, we first demonstrate how the regularization coefficient $\lambda$ affects differential immunity $\gamma$ in Fig. \ref{fig10}. In particular, if we reduce $\lambda$ to 0, which means that all member models are identical, i.e., EI-MTD’s dynamic scheduling is disabled, there are only 27.34\% accuracy under PGD attacks for EI-MTD. On the contrary, if we increase $\lambda$ to 1 by differential knowledge described in Section 4.2, EI-MTD achieves 55.68\% accuracy. In fact, increasing $\lambda$ means increasing the importance of diversity during distillation in Eq. (9), which accordingly increases the differential immunity. In this way, it shows larger $\gamma$ can improve the accuracy of EI-MTD. Fig. \ref{fig11} demonstrates that increasing $\gamma$ can improve the accuracy with the distillation temperature $T=10$. As a result, we briefly summarize the above analysis as follows: (1) larger $\lambda$ can increase the diversity of member models, which further improves the differential immunity $\gamma$, (2) larger $\gamma$ can guarantee higher accuracy of EI-MTD and (3) so larger $\lambda$ is of benefit to improve accuracy.

\begin{figure}   
\centering   
\subfigure[FGSM]{  \includegraphics[width=4cm,height=3cm]{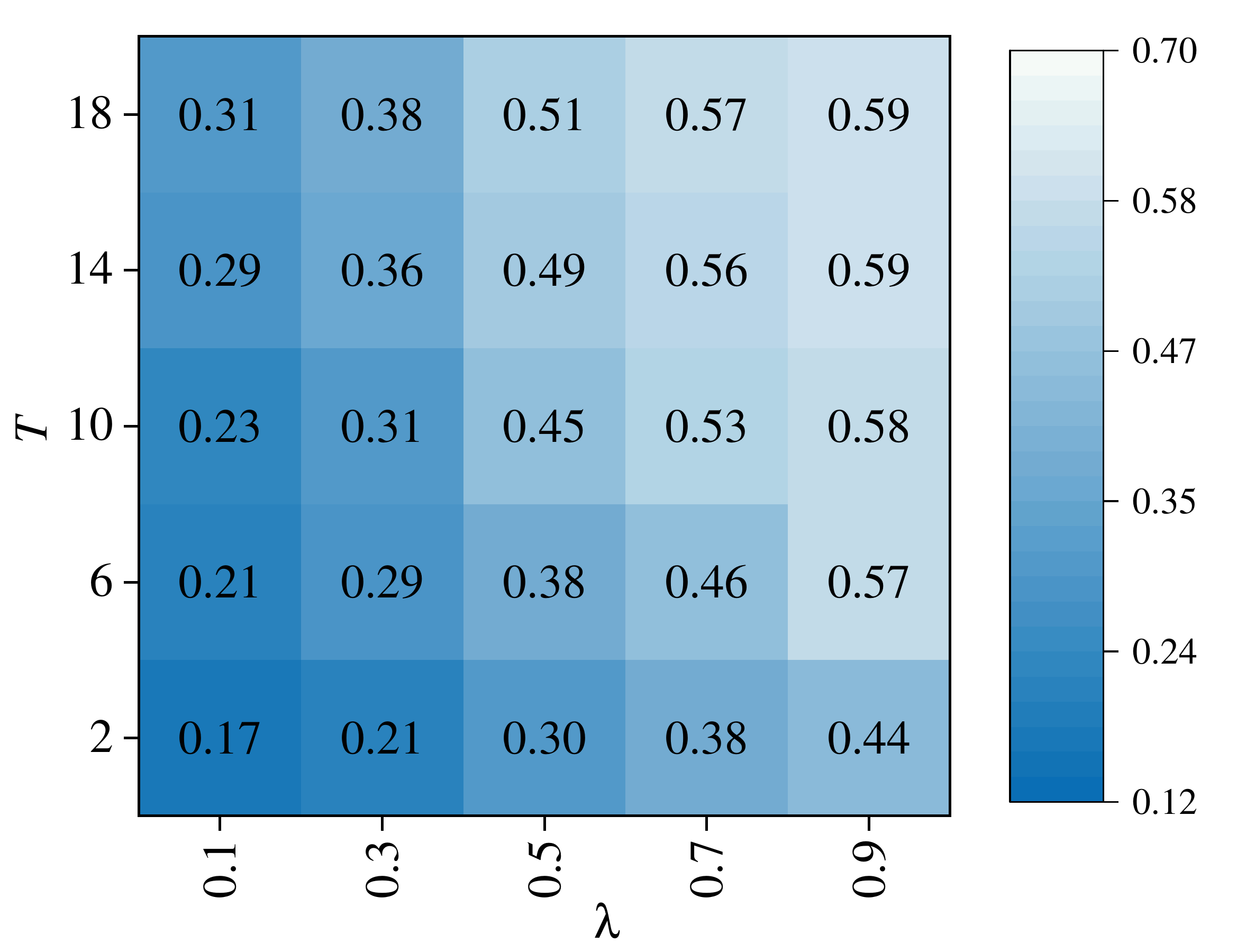}    \includegraphics[width=4cm,height=3cm]{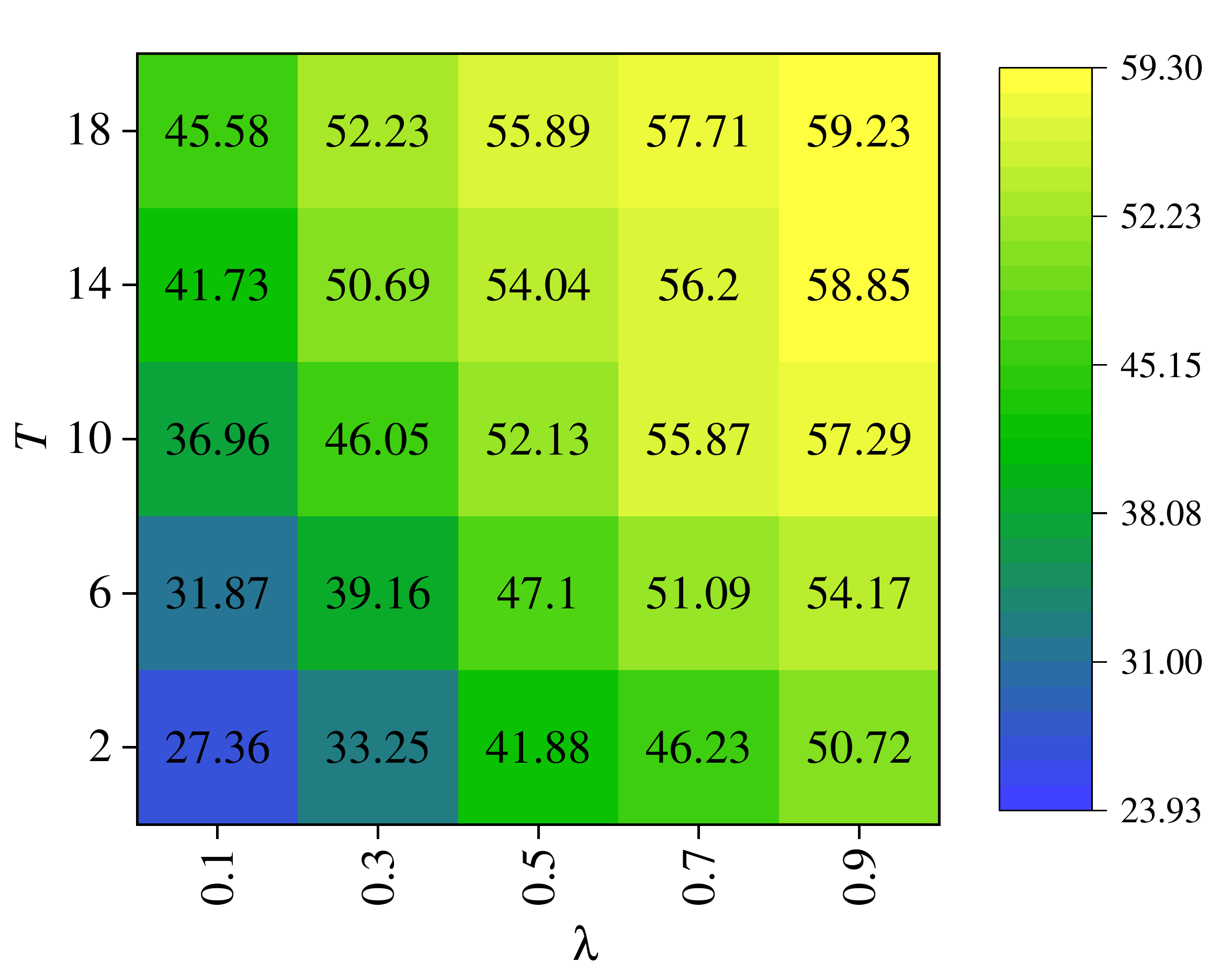}     }   

\subfigure[PGD]{\includegraphics[width=4cm,height=3cm]{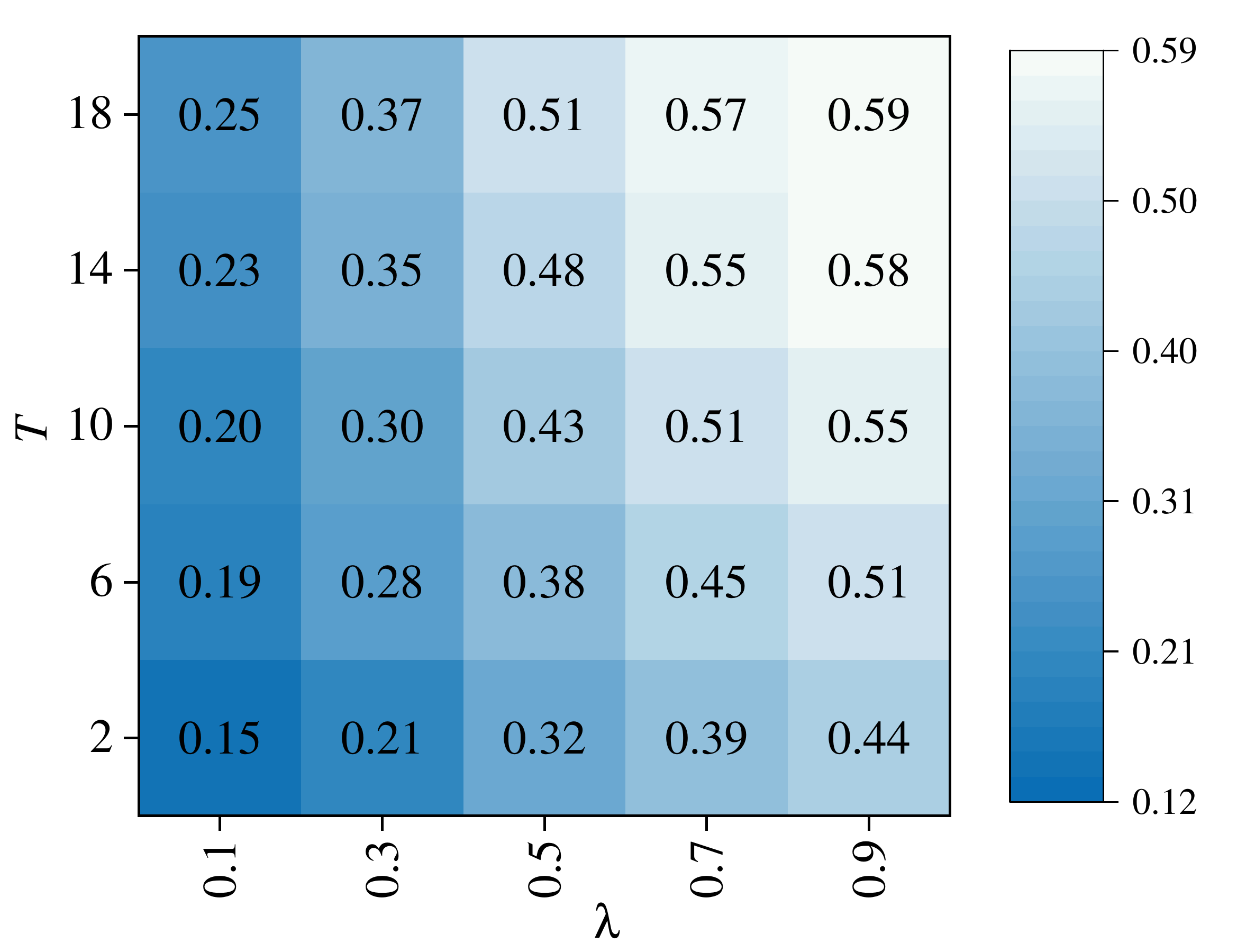}    \includegraphics[width=4cm,height=3cm]{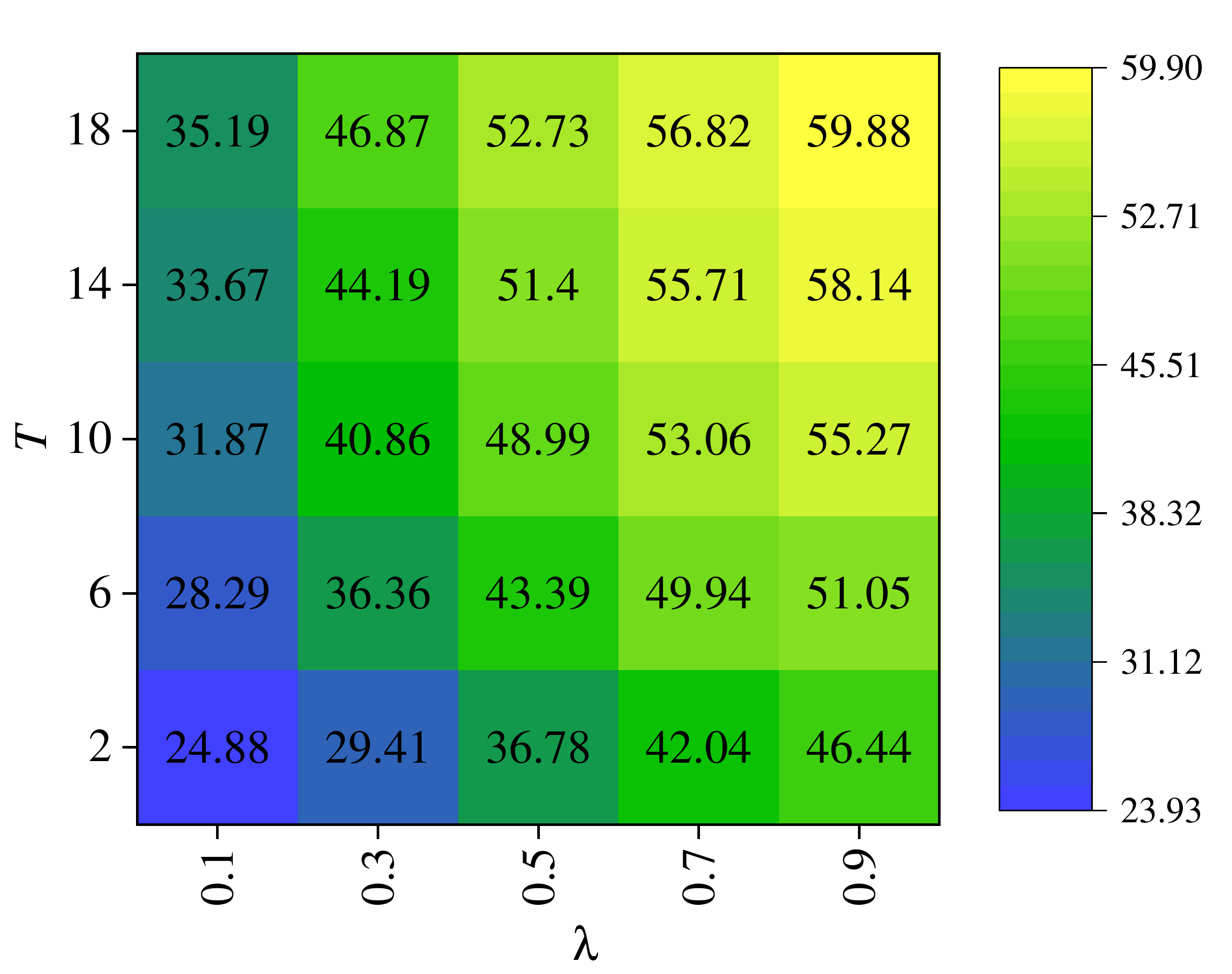}   }   

\subfigure[MI-FGSM]{\includegraphics[width=4cm,height=3cm]{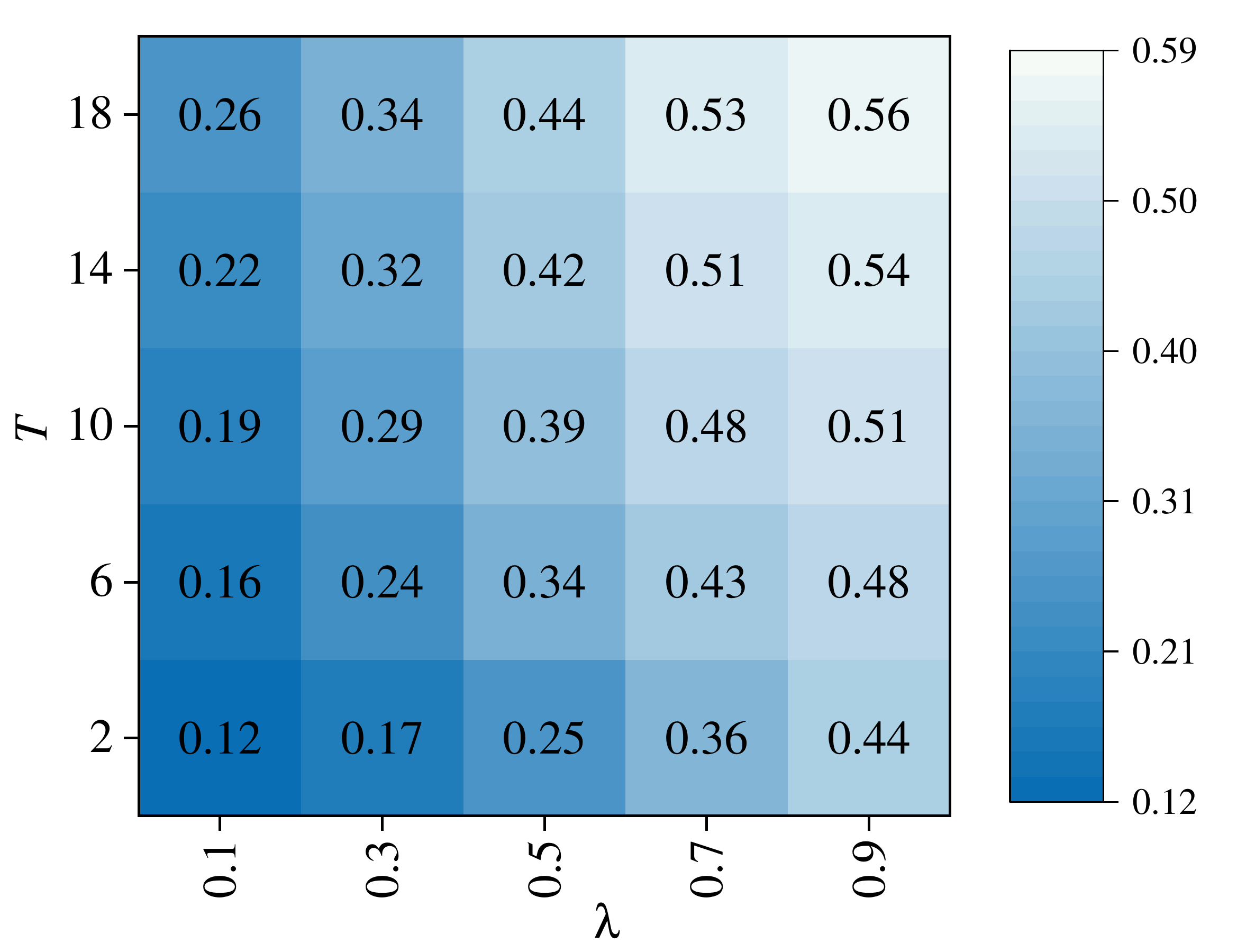}    \includegraphics[width=4cm,height=3cm]{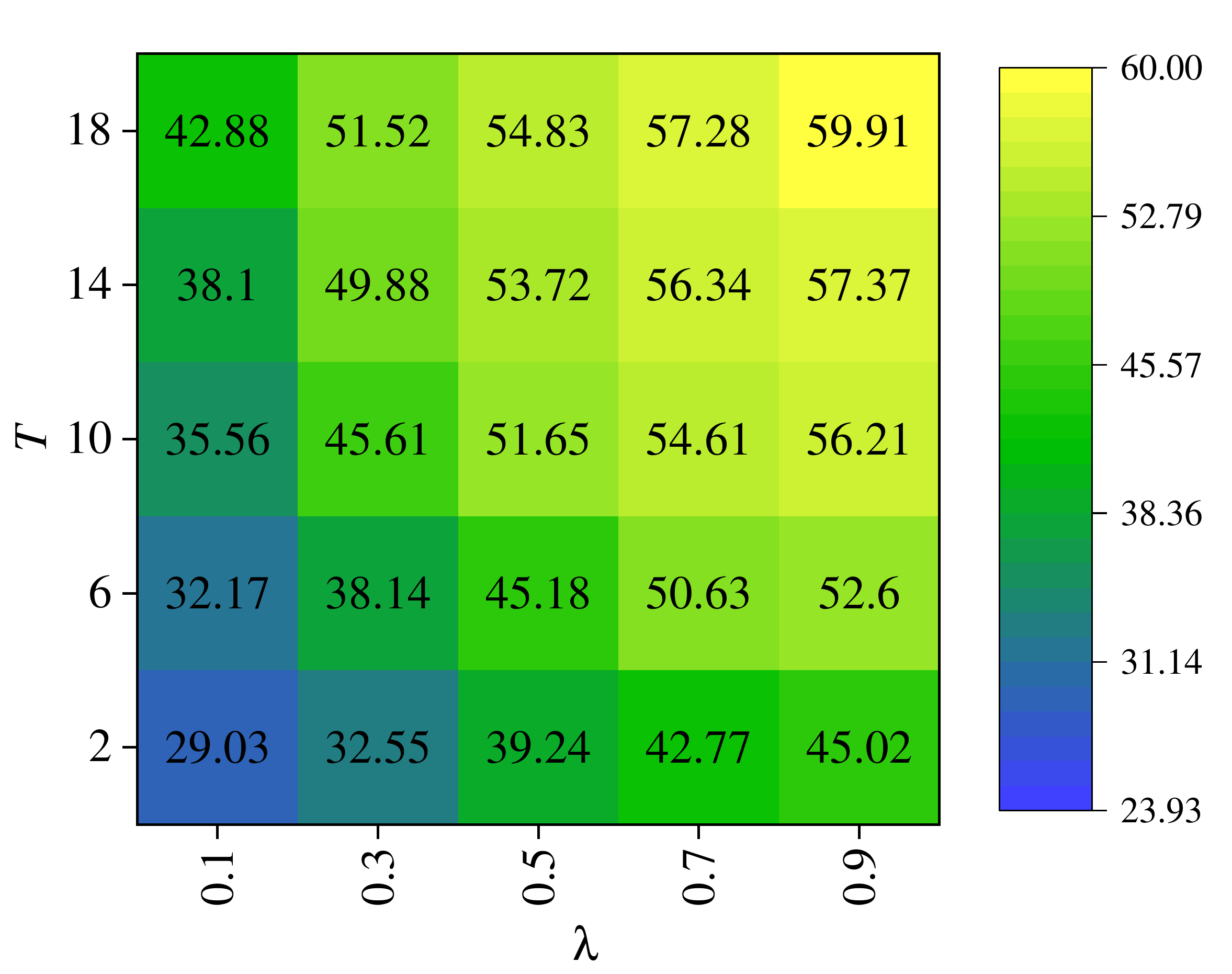}   }

\subfigure[M-DI$^{2}$-FGSM]{\includegraphics[width=4cm,height=3cm]{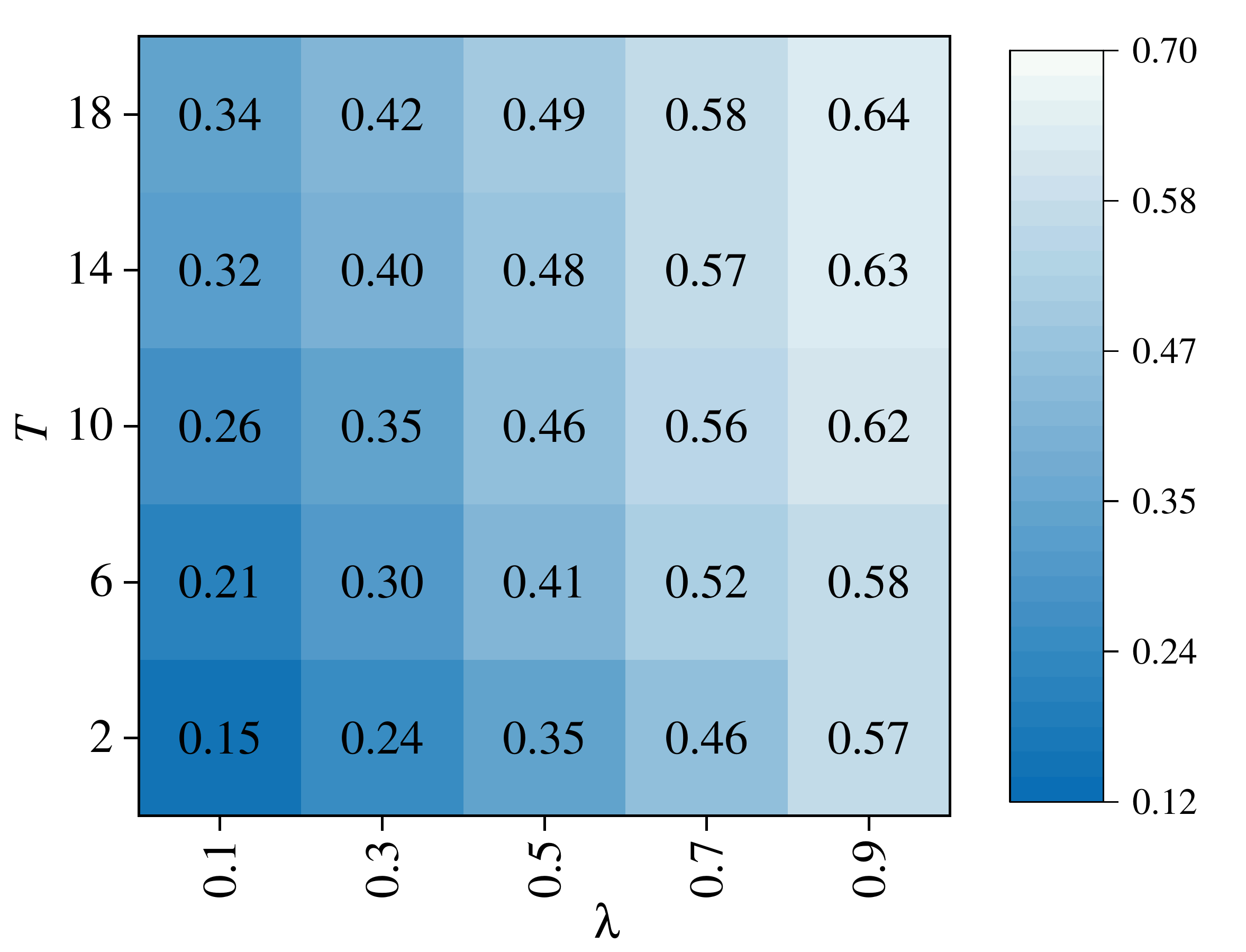}    \includegraphics[width=4cm,height=3cm]{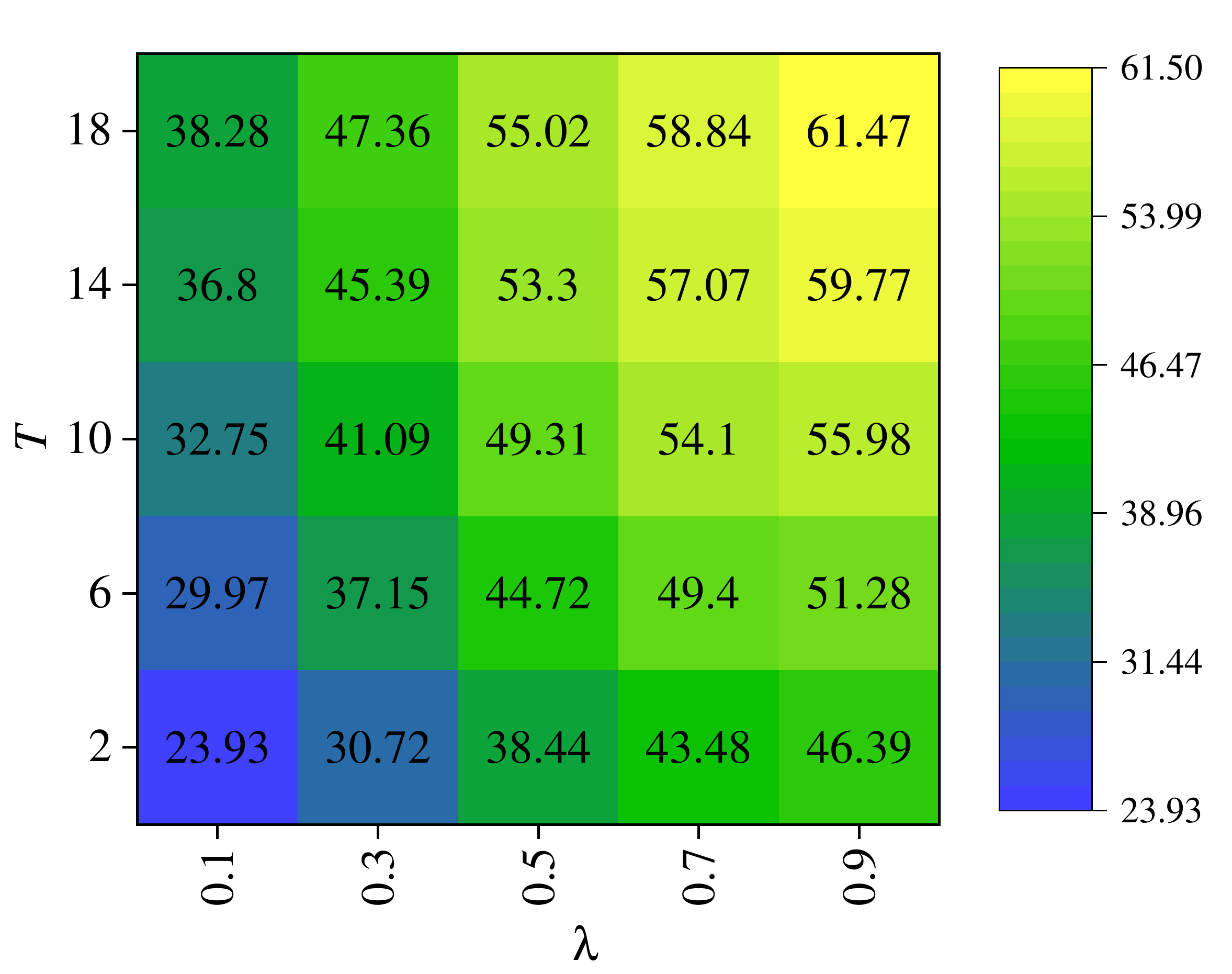}   }

\caption{The thermal diagram of the differential immunity and accuracy with respect to the combination of $T$ and $\lambda$. The left column represents the differential immunity, and the right column represents the accuracy. (a), (b), (c), and (d) illustrate the adversarial attacks including FGSM, PGD, MI-FGSM, and M-DI$^{2}$-FGSM respectively.}  
\label{fig12}   
\end{figure}

\textbf{Optimal combination of $T$ and $\lambda$.} Although we analyzed the impact of $T$ and $\lambda$ separately, it is not clear the impact of their combination on the accuracy of EI-MTD. We demonstrate the differential immunity and accuracy through the thermal diagrams in Fig. \ref{fig12}. Luckily, $T$ and $\lambda$ do not counteract each other’s impact on EI-MTD. Therefore, increasing $T$ and $\lambda$ at the same time can improve the accuracy of EI-MTD, which is similar to the differential immunity. In our experiments, $T=18$ and $\lambda=0.9$ can achieve a perfect performance, but too larger values seem to have no further significant effect.


\section{RELATED WORK}
Due to few works on adversarial attacks against EI, we mainly summarize the adversarial attacks and their countermeasures on the cloud data center.

\subsection{Adversarial Attacks}
An adversarial attack is to produce adversarial examples that lead to misclassiﬁcation when added some well-designed perturbations to the benign inputs. Recent literature has proposed multiple methods of crafting adversarial examples for DNNs \cite{[a14]}, \cite{[a15]}, \cite{[a21]}, \cite{[a41]}, \cite{[a82]}. The process of crafting adversarial examples is usually formalized as an optimization problem based on gradients. FGSM \cite{[a15]} generates adversarial examples with a single gradient step. Later, the basic iterative method called I-FGSM \cite{[a16]} improved upon FGSM was proposed, which takes multiple, smaller FGSM steps, ultimately rendering both FGSM-based adversarial training ineffective \cite{[a16]}. The PGD attack is considered as the strongest ﬁrst-order attack that uses first-order gradient descent to ﬁnd adversarial examples \cite{[a21]}. The methods to generate these adversarial examples often rely on the gradient of loss functions, such as FGSM \cite{[a15]}, I-FGSM \cite{[a16]}, and M-DI$^{2}$-FGSM \cite{[a17]}, etc.

Early on, researchers noticed that adversarial examples computed against one DNN are likely to be misclassiﬁed by other DNNs \cite{[a64]}, \cite{[a45]}. This phenomenon is termed transferability that serves as the basis of black-box attacks. The accepted explanation for the transferability phenomenon is that the gradients (which are used to compute adversarial examples) of each DNN are good approximators of those of other DNNs \cite{[a45]}, \cite{[a47]}. 

Previous methods assume adversarial data can be directly fed into deep neural networks. However, in many applications, people can only pass data through devices (e.g., cameras, sensors). Kurakin et al. applied adversarial examples to the physical world \cite{[a16]}. Eykholt et al. [a48] attack a vehicle vision system by modifying a stop sign as a speed limit sign. Sharif et al. \cite{[a49]} attack a state-of-the-art face-recognition algorithm through printing a pair of eyeglass frames. Hence, adversarial attacks become a real thread to edge intelligence. 

\subsection{Countermeasures}
A number of defenses have been proposed to harden networks against adversarial attacks, including preprocessing techniques \cite{[a50]}, \cite{[a51]}, detection algorithms \cite{[a53]}, \cite{[a54]}, veriﬁcation and provable defenses \cite{[a58]}, \cite{[a59]}, and various theoretically motivated heuristics \cite{[a60]}, \cite{[a61]}. More sophisticated defenses that rely on network distillation \cite{[a62]} and specialized activation functions \cite{[a63]} were also toppled by strong attacks \cite{[a64]}, \cite{[a65]}, \cite{[a41]}, \cite{[a63]}. Despite the eventual defeat of other adversarial defenses, adversarial training with a PGD adversary remains empirically robust to this day \cite{[a21]}. In \cite{[a21]}, it shows that increasing the capacity of DNNs can increase the robustness when adversarial training is applied. Hence, we choose adversarial training for a teacher DNN on the cloud data center, but not train student DNNs directly due to their relatively small capacity. Instead, we further use knowledge distillation to transfer robust knowledge from a teacher DNN to student DNNs.

Many recent defenses resort to randomization schemes for mitigating the effects of adversarial perturbations in the input or feature space. Xie et al. \cite{[a68]} utilize random resizing and padding to mitigate the adversarial effects at the inference stage. Guo et al. \cite{[a50]} apply image transformations with randomness such as a bit-depth reduction before feeding the image to a CNN. These methods are effective in black-box settings, but not well in white-box settings. Dhillon et al. \cite{[a69]} present a method to stochastically prune a subset of the activations in each layer and retain activations with larger magnitudes. Luo et al. \cite{[a70]} introduce a new CNN structure by randomly masking the feature maps output from the convolutional layers. 

Other randomization schemes are inspired by MTD, which randomly switches the models served for classification. Sailik et. al \cite{[a23]} proposed an MTDeep to defend DNNs by a repeated Bayesian game. Their experimental results show that MTDeep reduces misclassiﬁcation on perturbed images for MNIST and ImageNet datasets while maintaining high classiﬁcation accuracy on a legitimate test image. Roy et. al \cite{[a24]} propose an MTD against transferable adversarial attacks. Their experiment results show that even under very harsh constraints, e.g., no attack-cost, and availability of attacks which can bring down the accuracy to 0, it is possible to achieve reasonable accuracy for classification. Song et. al \cite{[a25]} designed an fMTD for embedded deep visual sensing systems by forking models to simulate MTD. However, all these defenses are not aimed at edge intelligence and did not take account of the limitation of resources as we explained in Section 1.

\section{CONCLUSION}
In this work, we focus on improving robustness, especially the classification accuracy of EI in adversarial settings. We propose EI-MTD to defend against adversarial attacks. Several techniques are integrated into our EI-MTD framework, including differential knowledge distillation and Bayesian Stackelberg games. The experiment results show that these techniques can effectively improve the robustness of edge intelligence against attacks. Especially, increasing distillation temperature $T$ and regularization coefficient $\lambda$ can effectively improve EI-MTD’s performance. In the future, we will explore an ensemble framework based on EI-MTD, which hopes to further improve the accuracy of adversarial examples.

\ifCLASSOPTIONcompsoc
  \section*{acknowledgments}
\else
  \section*{acknowledgment}
\fi

This work is supported by the Zhejiang Provincial Natural Science Foundation of China (No.LGF20F020007, No.LY18F020012), Natural Science Foundation of China (No.61902082, 61972357), National Key R\&D Program of China (No.2019YFB1706003) and Zhejiang Key R\&D Program (No.2019C03135).

\ifCLASSOPTIONcaptionsoff
  \newpage
\fi

\begin{IEEEbiography}[{\includegraphics[width=1in,height=1.25in,clip,keepaspectratio]{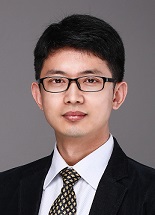}}]{Yaguan Qian} is an associate professor of School of Big-data Science, Zhejiang University of Science \& Technology. He received his B.S. in Computer Science from Tianjin University in 1999, and MS and Ph.D in Computer Science from Zhejiang University in 2005 and 2014 respectively. His research interests mainly include adversarial machine learning and edge computing. In these areas he has published more than 30 papers in international journals or conferences. He is the member of CCF (China Computer Federation), CAAI (Chinese Association for Artificial Intelligence) and CAA (Chinese Association of Automation).
\end{IEEEbiography}

\begin{IEEEbiography}[{\includegraphics[width=1in,height=1.25in,clip,keepaspectratio]{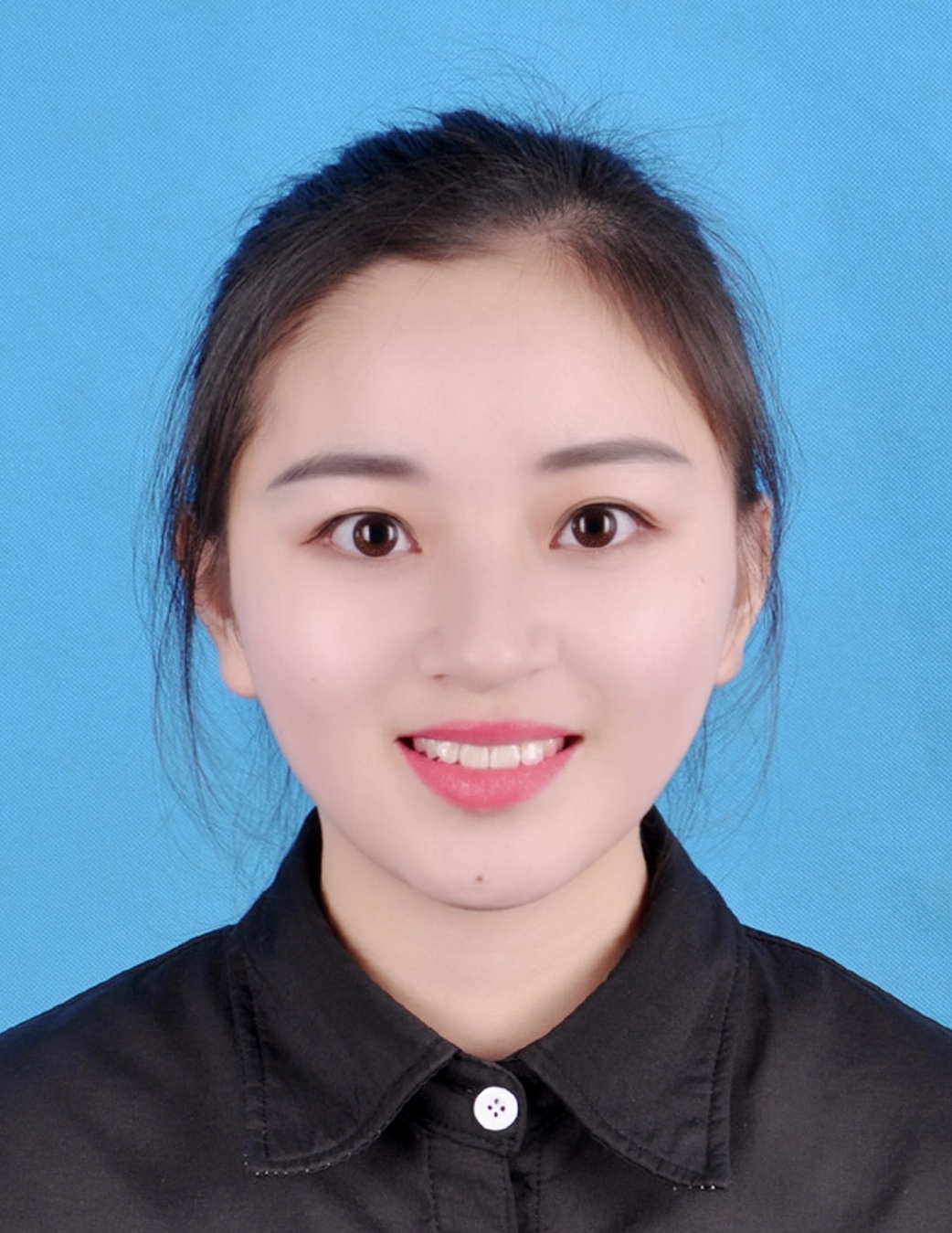}}]{Qiqi Shao} is currently a M.S student in the college of science at Zhejiang University of Science and Technology. Her current research interests include deep learning security and machine learning security.
\end{IEEEbiography}

\begin{IEEEbiography}[{\includegraphics[width=1in,height=1.25in,clip,keepaspectratio]{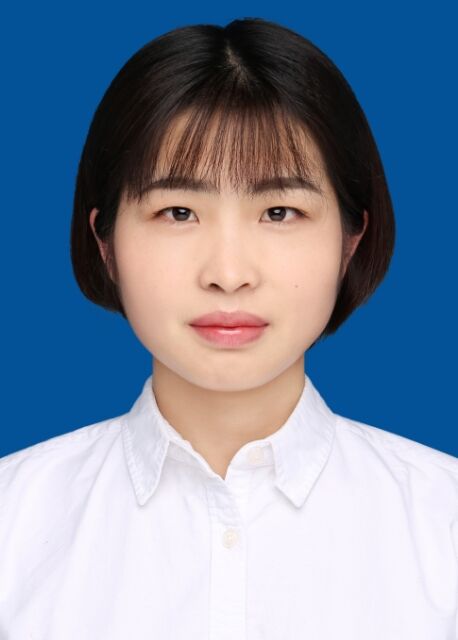}}]{Jiamin Wang} received the bachelor’s degree from Zhejiang Gongshang University Hangzhou College of Commerce in 2017. She is now a postgraduate student of the School of Big-data Science, Zhejiang University of Science and Technology. Her research interests include deep learning and machine learning security.
\end{IEEEbiography}

\begin{IEEEbiography}[{\includegraphics[width=1in,height=1.25in,clip,keepaspectratio]{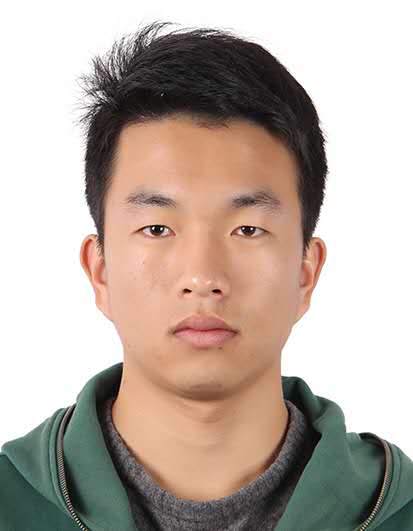}}]{Xiang Ling} is currently a Ph.D student in the College of Computer Science and Technology at Zhejiang University. His current research interests include AI security and data-driven security.
\end{IEEEbiography}

\begin{IEEEbiography}[{\includegraphics[width=1in,height=1.25in,clip,keepaspectratio]{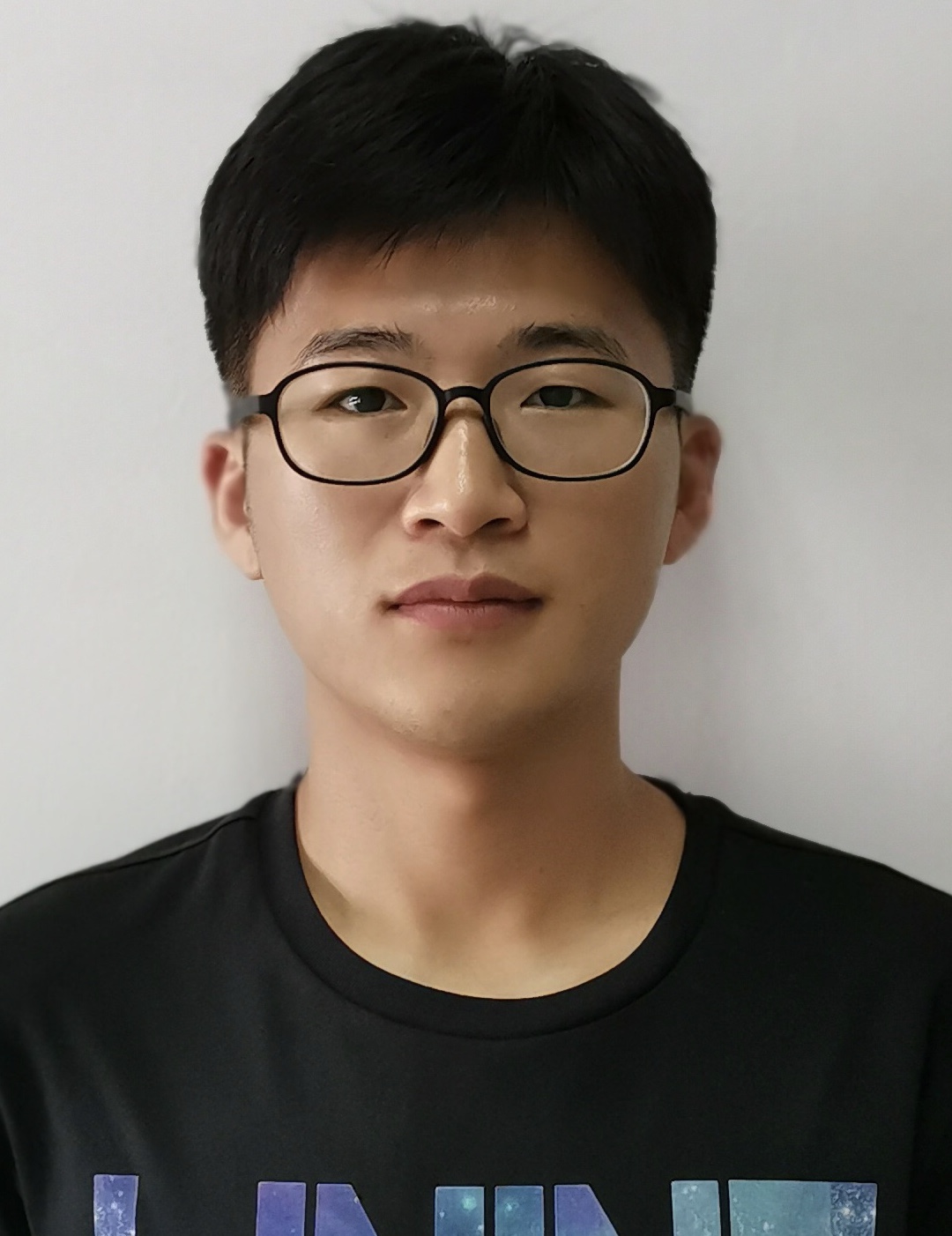}}]{Yankai Guo} is currently a M.S student in the college of science at Zhejiang University of Science and Technology.His current research interests include deep learning security and image recognition.
\end{IEEEbiography}

\begin{IEEEbiography}[{\includegraphics[width=1in,height=1.25in,clip,keepaspectratio]{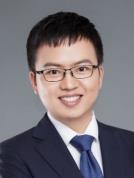}}]{Zhaoquan Gu} received his bachelor degree in Computer Science from Tsinghua University (2011) and PhD degree in Computer Science from Tsinghua University (2015). He is currently a professor in the Cyberspace Institute of Advanced Technology (CIAT), Guangzhou University, China. His research interests include wireless networks, distributed computing, big data analysis, and artificial intelligence security.
\end{IEEEbiography}

\begin{IEEEbiography}[{\includegraphics[width=1in,height=1.25in,clip,keepaspectratio]{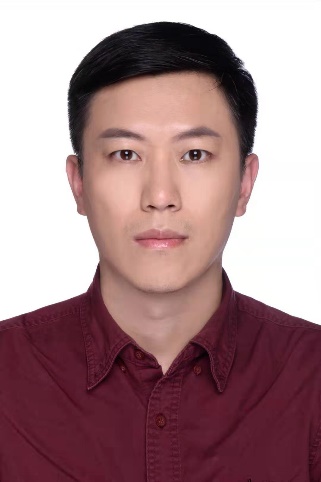}}]{Bin Wang}, Ph.D., Senior Engineer, he is the Vice President of Hangzhou Hikvision Digital Technology Co., Ltd. His research interests mainly include new computer networks, artificial intelligence security, Internet of Things security and so on. In these areas he has published more than 50 papers in international journals or conferences. He is the member of CCF (China Computer Federation) and CIEE Chinese Institute of Electronics
\end{IEEEbiography}

\begin{IEEEbiography}[{\includegraphics[width=1in,height=1.25in,clip,keepaspectratio]{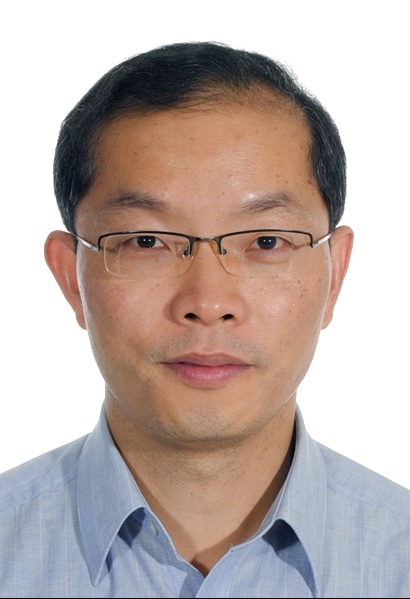}}]{Chunming Wu} received the Ph.D. degree in computer science from Zhejiang University in 1995. He is currently a Professor with the College of Computer Science and Technology, Zhejiang University, where he is also an Associate Director of Institute of Computer System Architecture and Network Security of Zhejiang University.  His research fields include software-defined networks, reconfigurable networks, proactive network defense, network virtualization and intelligent networks.
\end{IEEEbiography}

\begin{IEEEbiographynophoto}{   }

\end{IEEEbiographynophoto}

%

%
%
%




\end{document}